\documentclass[journal,onecolumn]{IEEEtran}
\ifdefined\XeTeXversion
\AtBeginDvi{}
\fi
\usepackage[T1]{fontenc}
\usepackage{amsmath,amssymb,amsfonts}
\usepackage{mathtools}
\usepackage[table]{xcolor}
\usepackage{array}
\usepackage{graphicx}
\usepackage{tikz}
\definecolor{assigned}{rgb}{0.15,0.32,0.55}
\newcommand{\nz}[1]{{\color{assigned}#1}}
\newcommand{\ind}[1]{\mathbf1_{\{#1\}}}
\newcommand{\ZZ}{\mathsf Z}
\newsavebox{\matrixbox}
\newcommand{\fitmatrix}[1]{\sbox{\matrixbox}{$\displaystyle #1$}%
\ifdim\wd\matrixbox>\linewidth\resizebox{\linewidth}{!}{\usebox{\matrixbox}}%
\else\usebox{\matrixbox}\fi}
\usetikzlibrary{matrix,fit,calc,arrows.meta,bending,positioning,arrows}
\usepackage{textcomp}
\usepackage{url}
\usepackage{cite}
\makeatletter
\newenvironment{inlinefigure}{\par\medskip\noindent
\begin{minipage}{\linewidth}\def\@captype{figure}\centering}
{\end{minipage}\par\medskip}
\makeatother

\definecolor{revisionblue}{rgb}{0.02,0.16,0.78}
\ifdefined\ShowReviewChanges
\newcommand{\revision}[1]{{\color{revisionblue}#1}}
\newenvironment{revisionblock}{\begingroup\color{revisionblue}%
\renewcommand{\nz}[1]{##1}}{\endgroup}
\else
\newcommand{\revision}[1]{#1}
\newenvironment{revisionblock}{}{}
\fi

\ifdefined\ShowSRNames
\newcommand{\refine}[1]{#1}
\newenvironment{refineblock}{}{}
\newcommand{\arxivfix}[1]{#1}
\newcommand{\srnamefix}[1]{{\color{revisionblue}#1}}
\else
\newcommand{\refine}[1]{#1}
\newenvironment{refineblock}{}{}
\newcommand{\arxivfix}[1]{#1}
\newcommand{\srnamefix}[1]{#1}
\fi

\ifdefined\ShowSecondReview
\newcommand{\reviewfix}[1]{{\color{revisionblue}#1}}
\newenvironment{reviewblock}{\begingroup\color{revisionblue}}{\endgroup}
\else
\newcommand{\reviewfix}[1]{#1}
\newenvironment{reviewblock}{}{}
\fi

\ifdefined\ShowThirdReview
\newcommand{\thirdfix}[1]{{\color{revisionblue}#1}}
\newenvironment{thirdblock}{\begingroup\color{revisionblue}}{\endgroup}
\else
\newcommand{\thirdfix}[1]{#1}
\newenvironment{thirdblock}{}{}
\fi

\ifdefined\ShowFourthReview
\newcommand{\fourthfix}[1]{{\color{revisionblue}#1}}
\newenvironment{fourthblock}{\begingroup\color{revisionblue}}{\endgroup}
\else
\newcommand{\fourthfix}[1]{#1}

\fi

\ifdefined\ShowFifthReview
\newcommand{\fifthfix}[1]{{\color{revisionblue}#1}}
\newenvironment{fifthblock}{\begingroup\color{revisionblue}}{\endgroup}
\else
\newcommand{\fifthfix}[1]{#1}

\fi

\ifdefined\ShowSixthReview
\newcommand{\sixthfix}[1]{{\color{revisionblue}#1}}
\newenvironment{sixthblock}{\begingroup\color{revisionblue}}{\endgroup}
\else
\newcommand{\sixthfix}[1]{#1}

\fi

\ifdefined\ShowSeventhReview
\newcommand{\seventhfix}[1]{{\color{revisionblue}#1}}
\else
\newcommand{\seventhfix}[1]{#1}
\fi

\ifdefined\ShowEighthReview
\newcommand{\eighthfix}[1]{{\color{revisionblue}#1}}
\newenvironment{eighthblock}{\begingroup\color{revisionblue}}{\endgroup}
\else
\newcommand{\eighthfix}[1]{#1}
\newenvironment{eighthblock}{}{}
\fi

\newtheorem{theorem}{Theorem}         
\newtheorem{lemma}[theorem]{Lemma}             
\newtheorem{definition}{Definition}    
\newtheorem{example}{Example}          
\newtheorem{corollary}[theorem]{Corollary}      
\newtheorem{remark}{Remark}           

\newtheorem{proposition}[theorem]{Proposition}
\newsavebox{\fitgraphicbox}
\newcommand{\fitgraphic}[2]{\sbox{\fitgraphicbox}{#2}%
\ifdim\wd\fitgraphicbox>#1\resizebox{#1}{!}{\usebox{\fitgraphicbox}}%
\else\usebox{\fitgraphicbox}\fi}
\begin{document}
\title{\srnamefix{Rate-Optimal Streaming Codes Under an Extended Delay Profile for Three-Node Relay Networks With Burst Erasures}}
\author{Ke Wang\textsuperscript{\textdaggerdbl}, Wenjie Ma\textsuperscript{\textsection},
Zhipeng Li\textsuperscript{\textdagger,*}, and Yaqiang Zhang\textsuperscript{\textdaggerdbl}%
\thanks{\raggedright\textsuperscript{\textdagger}Zhipeng Li is with the School of Computing and Information Science,
Fuzhou Institute of Technology, Fuzhou 350506, China
(e-mail: \texttt{lzp13484@gmiot.com}).}%
\thanks{\raggedright\textsuperscript{\textdaggerdbl}Ke Wang and Yaqiang Zhang are with IEIT SYSTEMS Co., Ltd.,
Jinan 250101, China
(e-mail: \texttt{wangke12@ieisystem.com}; \texttt{zhangyaqiang@ieisystem.com}).}%
\thanks{\raggedright\textsuperscript{\textsection}Wenjie Ma is with the School of Mathematics,
Southwest Minzu University, Chengdu 610225, China
(e-mail: \texttt{mawenjie@swun.edu.cn}).}%
\thanks{\raggedright\textsuperscript{*}Corresponding author: Zhipeng Li.}}
\maketitle
\ifdefined\ShowSecondReview
\begin{center}\footnotesize
\reviewfix{Blue text marks the changes made in this review round.}
Shaded matrix entries retain their construction-specific meaning.
\end{center}
\vspace{0.6\baselineskip}
\fi

\ifdefined\ShowThirdReview
\begin{center}\footnotesize
\thirdfix{Blue text marks the changes in the third review round.}
Shaded matrix entries retain their construction-specific meaning.
\end{center}
\vspace{0.6\baselineskip}
\fi

\ifdefined\ShowFourthReview
\begin{center}\footnotesize
\fourthfix{Blue text marks the changes in the fourth review round.}
Shaded matrix entries retain their construction-specific meaning.
\end{center}
\vspace{0.6\baselineskip}
\fi

\ifdefined\ShowFifthReview
\begin{center}\footnotesize
\fifthfix{Blue text marks the prior-work and table-reference update.}
Shaded matrix entries retain their construction-specific meaning.
\end{center}
\vspace{0.6\baselineskip}
\fi

\ifdefined\ShowSixthReview
\begin{center}\footnotesize
\sixthfix{Blue text marks the revised introduction structure and table framing.}
Shaded matrix entries retain their construction-specific meaning.
\end{center}
\vspace{0.6\baselineskip}
\fi

\ifdefined\ShowSeventhReview
\begin{center}\footnotesize
\seventhfix{Blue text marks the worked-example transmission update.}
Red terms in the transmission table belong to the source packet $S[0]$.
\end{center}
\vspace{0.6\baselineskip}
\fi

\ifdefined\ShowEighthReview
\begin{center}\footnotesize
\eighthfix{Blue text marks the main-text proofs and their revised transitions.}
\end{center}
\vspace{0.6\baselineskip}
\fi

\begin{abstract}
We study streaming codes for a three-node relay network with burst lengths $b_1,b_2$ and an end-to-end decoding delay $T$. Let $u=\max\{b_1,b_2\}$ and $v=\min\{b_1,b_2\}$, and write $T-v=pu+q$, where $T\ge u+v$ and $0\le q<u$. We develop two binary constructions, SR and SR-2, using extended delay profiles that allow the relay to forward causally invertible combinations of current and past source symbols. The SR construction attains the non-adaptive rate upper bound whenever
\[
\frac{T-u-v}{2u-v}\le\left\lfloor\frac{T-u-v}{u}\right\rfloor.
\]
SR-2 enlarges this range under the sufficient condition
\[
q(v-q)\le(p-1)u(u-v).
\]
These constructions complement the known divisibility and large-delay regimes. We also establish two limitations of systematic ordinary coordinate forwarding. First, it cannot attain the rate upper bound when $0<T-u-v<v$. Second, for equal burst lengths $b$, it attains the bound if and only if $b\mid(T-2b)$; a quantitative upper bound gives a strictly positive rate gap in every nondivisible case. These converses allow nonlinear components and predetermined time variation within the stated ordinary-forwarding architecture. Existing adaptive schemes for equal bursts approach the same rate upper bound asymptotically using time-varying convolutional codes and lie outside the non-adaptive setting considered here.
\end{abstract}
\newenvironment{proof}[1][Proof]{
		\par                            
		\normalfont
		\topsep6pt                       
		\trivlist                        
		\item[\hskip1em\hskip\labelsep\textit{#1:}] 
		\ignorespaces                        
	}{%
		\nobreak\hfill$\square$          
		\endtrivlist                    
	}

\section{Introduction}
\IEEEPARstart{S}{treaming} codes protect sequential messages under
strict decoding deadlines. Martinian and Sundberg introduced
point-to-point codes for burst erasures \cite{1337126}.
Subsequent work established the capacity of sliding-window channels
with burst and arbitrary erasures and developed explicit
low-delay constructions over smaller fields
\cite{7593328,9047134,8621051,9578987,8917664}.
Delay-optimal and low-complexity designs were also studied
\cite{4557355,6567095,6620379,7839973,9611442}.

In a three-node network, the source reaches the destination through
a relay. The source-to-relay (SR) and relay-to-destination (RD)
links may experience independent burst erasures, while every source
packet must be recovered within a single end-to-end deadline.
A coordinate recovered later at the relay leaves less time for
protection on the RD link.

Three-node streaming codes for arbitrary erasures were studied
in \cite{8835153,10064107}. Singhvi et al. derived the non-adaptive
rate bound \cite{9834645}
\begin{equation}\label{ob}
\min\left\{\frac{T-b_1-a_2+1}{T-b_1+b_2-a_2+1},
\frac{T-b_2-a_1+1}{T-b_2+b_1-a_1+1}\right\}
\end{equation}
when link $i$ permits a burst of length $b_i$ or $a_i$ arbitrary
erasures. For burst-only channels, setting $a_1=a_2=1$ gives
\[
R(b_1,b_2,T)=
\min\left\{\frac{T-b_1}{T-b_1+b_2},
\frac{T-b_2}{T-b_2+b_1}\right\}.
\]
Their construction attains this bound when
$\max\{b_1,b_2\}\mid(T-b_1-b_2)$.
For equal bursts, \cite{10764771} gives asymptotically optimal
codes whose relay information sets adapt to the SR burst.
For unequal bursts, Li et al.~\cite{li2025rateoptimalstreamingcodesthreenode}
constructed binary ordinary coordinate-forwarding codes that attain
this bound when $T\ge b_1+b_2+b_1b_2/|b_1-b_2|$.
Related multihop and multi-access models appear in
\cite{9814088,9916294}. Recent work also studies adaptive burst-erasure
relaying~\cite{JayakrishnanAdaptive2026}. The construction coverage
compared below concerns exact attainment by fixed-rate non-adaptive
codes; adaptive results are a separate comparison because their
relay output may depend on the SR erasure history.

A code is non-adaptive if, for a fixed source stream, its transmitted
relay packets are independent of the SR erasure history. This distinction is also made in the correction
\cite{FongCorrection2022}. Our constructions pair point-to-point
components, and Section~\ref{sec:ordinary-forwarding} studies the specific case in which the
relay forwards original coordinates with fixed delays.

\subsection{\sixthfix{Main Results}}
\label{sec:main-overview}
Write $T-v=pu+q$ and put $A=(p-1)(u-v)$.
Our main results give two binary constructions, SR and SR-2, that
attain the known non-adaptive rate upper bound~\cite{9834645}, together
with two converses for systematic ordinary coordinate forwarding.
The nondivisible achievability results assume $u>v$, $p\ge2$,
and $0<q<u$; the divisible case is treated separately.

\begin{enumerate}
	\item \emph{Binary exact attainment.}
	Both constructions use $(k,n)=(T-v,T-v+u)$ and attain
	\[
	R^\star=\frac{T-v}{T-v+u}.
	\]
	The SR construction achieves $R^\star$ whenever
	\[
	\frac{T - u - v}{2u - v} \le \left\lfloor\frac{T - u - v}{u}\right\rfloor,
	\]
	while the SR-2 construction achieves $R^\star$ under the weaker sufficient condition
	\[
	q(v-q)\le u(p-1)(u-v),
	\]
	for either link order (Theorem~\ref{pw:achievability}).
\item \emph{Impossibility for ordinary forwarding.}
Systematic ordinary coordinate forwarding cannot attain $R^\star$
when $0<T-u-v<v$. This converse permits nonlinear components and
predetermined time variation (Theorem~\ref{secvi:theorem20}).
	\item \emph{Quantitative equal-burst impossibility bound.}
	For $u=v=b$ and $T-b=pb+q$, every ordinary code in Definition~\ref{def:ordinary-class},
	including nonlinear predetermined time-varying components, obeys
	\[
	R\le R_{\rm ord}=\frac{bp(p+1)}{b(p+1)^2-q}.
	\]
	When $q>0$, $R_{\rm ord}<R^\star$, so this entire ordinary-forwarding class has a strictly positive rate gap.
	Tightness of this upper bound for nondivisible cases remains open (Theorem~\ref{thm:equal-ordinary-allp}).
\end{enumerate}

\begin{table*}[!ht]
	\centering\small
	\caption{\sixthfix{Construction ranges for $T-v=pu+q$, $0\le q<u$}}
	\label{tab:comparison}
	\fitgraphic{\textwidth}{\begin{tabular}{|l|c|c|c|}\hline
			Construction&Parameter condition&Relay operation&Attainment\\\hline
			Singhvi et al.~\cite{9834645}&$q=0$&Non-adaptive, ordinary&Exact\\\hline
			Ramkumar et al.~\cite{10764771}&$u=v$&Adaptive&Asymptotic\\\hline
			Li et al.~\cite{li2025rateoptimalstreamingcodesthreenode}
			&$T-u\ge v\lceil(T-v)/u\rceil$, $u>v$&Non-adaptive, ordinary&Exact\\\hline
			SR (Section III)&$q\le (p-1)(u-v)$&Non-adaptive, extended&Exact, binary\\\hline
			SR-2 (Section IV)&$q(v-q)\le(p-1)u(u-v)$&Non-adaptive, extended&Exact, binary\\\hline
	\end{tabular}}
	\par\smallskip
	\begin{minipage}{0.97\textwidth}\footnotesize
		The nondivisible SR and SR-2 rows assume $u>v$, $p\ge2$, and
		$0<q<u$. The divisible case $q=0$ is handled separately by ordinary
		identity taps. Exact attainment uses fixed finite packet dimensions.
		The SR condition $q \le (p-1)(u-v)$ is equivalent to
	\(
	\frac{T - u - v}{2u - v} \le \left\lfloor\frac{T - u - v}{u}\right\rfloor,
	\) and implies the SR-2 condition
		$q(v-q)\le(p-1)u(u-v)$. SR-2 admits additional parameter triples;
		the layered baseline is retained in the overall coverage comparison.
		\revision{Adaptive burst-erasure relaying is also studied
			in~\cite{JayakrishnanAdaptive2026}. Adaptive schemes are outside
			the exact non-adaptive coverage counts; the equal-burst adaptive
			row is included as a separate benchmark.}
	\end{minipage}
\end{table*}

\noindent\emph{A small example of the mechanism.}
At $(b_1,b_2,T)=(5,6,20)$, one has $(p,q,A)=(2,3,1)$
and $(k,n)=(15,21)$. The short-hop groups
$(F,A^{\rm o},B,U)$ have sizes $(3,3,3,6)$ and delays
$(14,11,11,5)$, where the last entry concerns an innovation.
The long-hop ordinary delays are $(6,9,9,15)$, so every paired
sum is $20$. In particular the relay forwards
\[
I_U[t]=\bigl(U_F[t]+B[t-5],\ U_A[t]+B[t-1]\bigr)
\]
at delay $5$. The destination receives the older $B$ symbols
in time to remove them. Here $q>A$ and $q(v-q)=uA$, so the example demonstrates
the enlargement provided by SR-2 beyond the SR range.
\reviewfix{This example uses only the constant coefficient $C_0=I_3$.}
Section~\ref{ex:sr2-56} gives the encoder and all recovery steps.

The SR and SR-2 constructions use a common ordinary RD component.
\reviewfix{The polynomial zigzag matrix family and its degree properties
	come from~\cite{DeTeranZigzag2016}. In SR-2, these properties establish
	full row rank for the actual finite tail window. Recovery from that
	window must then meet every coordinate deadline and supply a causal,
	erasure-independent relay map for both link orders.
	Section~\ref{sec:SR2construction} separates these steps.}

Section~\ref{sec:final-coverage} gives an infinite family in the
additional SR-2 range, reports finite coverage counts, and compares
the attainment ranges with the two ordinary-forwarding converse regions.

\subsection{Organization}
Section~\ref{sec:problem-setup} defines the model and states the binary
achievability result. Section~\ref{KEY CONSTRUCTION} presents SR and
the common RD component. Section~\ref{sec:SR2construction} develops
SR-2, proves its extended delay profile, and gives two examples.
Section~\ref{sec:large-delay} compares construction ranges and
ordinary-forwarding limitations, and Section~\ref{sec:examples}
gives further composition examples. Section~\ref{sec:ordinary-forwarding}
proves the low-delay and equal-burst converses.
Section~\ref{sec:conclusion} concludes the paper. The appendices provide
the recovery-block entries, detailed decoding and relay maps,
supporting point-to-point lemmas, and an ordinary code outside the
layered baseline range.

\section{Problem Setup}\label{sec:problem-setup}
\subsection{Notation}
	In this paper, $\mathbb{N}$ denotes the set of nonnegative integers, and $\mathbb{N}_+$ denotes the set of positive integers. For $i,j\in\mathbb{N}$, let $[i:j]=\{k\in\mathbb{N}:i\leq k\leq j\}$ and $[i]=[1:i]$. An interval is empty when its lower endpoint exceeds its upper endpoint.
	$I_n$ denotes the $n \times n$ identity matrix. 
	The spaces of length-$k$ row vectors and $k\times n$ matrices over $\mathbb{F}$ are denoted by $\mathbb{F}^k$ and $\mathbb{F}^{k\times n}$, respectively. For $P\in\mathbb{F}^{k\times n}$, $P(i,j)$ is its $(i,j)$ entry. Let $\{B_\ell:\ell\in S\}$ be $s\times t$ matrices and let $h:[a]\times[b]\to S$. We write $B=(B_{h(i,j)})_{i\in[a],j\in[b]}$ for the $sa\times tb$ block matrix whose $(i,j)$ block is $B_{h(i,j)}$. For $m\in\mathbb{N}$ and $n\in\mathbb{N}_+$, define the least positive residue $\langle m\rangle_n=1+((m-1)\bmod n)\in[n]$.
	Throughout, $\mathbb F$ is a finite field. All explicit constructions
	in this paper use the binary field $\mathbb{F}_2$.
\subsection{Point-to-Point Model}\label{sec:component-checks}
Let $b,T\in\mathbb{N}_+$. A point-to-point \emph{streaming code}
(SC) with parameters $(b,T)$ maps a causal source stream
$\{S[t]\}_{t\geq0}$ to a transmitted stream $\{X[t]\}_{t\geq0}$.
Within every window of $T+1$ time instants, the erased packets,
if any, form one contiguous burst of length at most $b$. The decoder
must recover $S[t]$ by $t+T$ for every $t$.

Our constructed component encoders are systematic, time-invariant
convolutional codes \cite{MaISIT25}. At time $t$, a
systematic encoder transmits $X[t]=(S[t],P[t])\in\mathbb{F}^n$,
where $S[t]\in\mathbb{F}^k$ and $P[t]\in\mathbb{F}^{n-k}$. With
generator matrices $G_i\in\mathbb{F}^{k\times n}$, the truncated
input-output relation is
\begin{align}
	&(X[0], X[1], \dots, X[t])\notag\\
	&=(S[0], S[1], \dots, S[t])\begin{pmatrix}
		G_0 & G_1 & \cdots & G_t\\
		& G_0 & \cdots & G_{t-1}\\
		\mathbf{0}  &  & \ddots & \vdots\\
		& & & G_0
	\end{pmatrix}\label{eq-encoding}\\
	&=(S[0], S[1], \dots, S[t])( G_{j-i} )_{i,j\in [t+1]} \;.\notag
\end{align}
We use zero initialization, i.e., $S[t]=\mathbf{0}$ for $t<0$.

For the finite-memory systematic components constructed below, the
generator matrices have the form
\begin{equation*}
	G_i=\begin{cases}
		(I_k\mid P_0), & i=0\\
		(\mathbf{0}_{k\times k}\mid P_i), & i\neq0,
	\end{cases}
\end{equation*}
where $P_i\in\mathbb F^{k\times(n-k)}$ and we take $P_i=\mathbf{0}$
for $i\notin[0:T]$. Consequently,
\begin{equation}\label{eq-parity}
	P[t]=\sum_{i=0}^T S[t-i]P_i\;.
\end{equation}
The code rate is defined as $\frac{k}{n}$.

For an isolated burst erasing $X[t],\ldots,X[t+b-1]$, subtract
the contributions of unerased source packets from the parity
packets received at times $t+b,\ldots,t+b+T-1$. The resulting
equations are
\begin{equation}\label{eq-recovery}
	\begin{split}
		(\bar{P}[t+b], \bar{P}[t+b+1], \dots, \bar{P}[t+b+T-1]) \\
		= \left( S[t], S[t+1], \dots, S[t+b-1] \right)\mathbf{P},
	\end{split}
\end{equation}
where $\bar P[i]$ denotes the parity packet after subtracting all known contributions, and
\begin{equation}\label{eq-matrixP}
	\mathbf{P} = 
	\begin{pmatrix}
		P_{b} & P_{b+1} & \ldots & P_{T} & 0& \cdots & 0 \\
		P_{b-1} & P_{b} & \cdots & P_{T-1} & P_{T} & \cdots &0    \\
		\vdots & \vdots & \vdots & \vdots & \vdots & \ddots & \vdots  \\
		P_1 & P_2 & \cdots & P_{T-b+1} & P_{T-b+2} & \cdots & P_T
	\end{pmatrix}
\end{equation}

Different coordinates of $S[t]=(S_1[t],\ldots,S_k[t])$ may have different recovery deadlines.

\noindent\emph{Burst separation and all-time checks.}
A component guarantee verified for every isolated burst extends to
the sliding-window model. If an earlier burst ends at $e$ and the
next begins at $c$, then $c\geq e+D+1$ for a full delay $D$.
All earlier erased inputs are therefore decoded before $c$, so the
decoder can reconstruct the missing packets and apply the same
isolated-burst decoder. Shorter bursts are covered by ignoring
extra received packets.

For an extended innovation with deadline $\delta\geq b$, the
corresponding parity packet may itself be erased. In that case its
burst cannot erase the systematic packet at the innovation's source
time. The source symbol and the required earlier inputs are then
available. This received-or-erased check will be used explicitly
when establishing the extended profile below.
\begin{definition}\label{def-delay-profile}
	A \emph{delay profile} of a $(b,T)$ SC is a vector $(d_1,\ldots,d_k)\in\mathbb{N}^k$ such that, for every $t\in\mathbb{N}$ and $i\in[k]$, $S_i[t]$ is guaranteed to be recovered by time $t+d_i$. Thus $d_i\leq T$ for every $i$.
These entries are guaranteed upper bounds, not necessarily the
smallest possible delays. A \emph{minimum guaranteed delay} is
expressly identified when needed, as in Lemma~\ref{secvi:late-coordinates}.
\end{definition}

\subsection{Three-Node Relay Model}\label{key}
A $(b_1,b_2,T)$ three-node burst-erasure-correcting streaming code
(TBSC) communicates from source $\mathsf{s}$ to destination
$\mathsf{d}$ through relay $\mathsf{r}$.
In every window of $T+1$ time instants, the SR and RD links may
independently erase one burst of at most $b_1$ and $b_2$ packets.
Every source packet must reach the destination within delay $T$.
Within slot $t$, the relay first receives the current SR packet
(or its erasure indicator), then forms and sends $Z[t]$; the
destination receives the current RD packet before updating its
decoder. Same-slot receive-then-send processing adds no extra
unit of delay.

\begin{lemma}\label{lemma:relay-decoding}
Every valid causal TBSC allows the relay to recover $S[t]$
by $t+T-b_2$.
\end{lemma}
\begin{proof}
Fix any admissible SR history, and erase the final $b_2$ RD
packets at $t+T-b_2+1,\ldots,t+T$.
The destination must recover $S[t]$ using relay packets sent
through $t+T-b_2$. The relay has already generated these
packets at that time, and can run the stipulated destination
decoder itself. Hence it can recover $S[t]$ by $t+T-b_2$.
\end{proof}

Our constructions use the following component architecture.
\begin{itemize}
\item The SR encoder transmits $X[t]=(S[t],P[t])\in\mathbb F^n$
and provides a $(b_1,T-b_2)$ point-to-point guarantee.
The relay receives $Y^{(s)}[t]=X[t]$ or an erasure symbol.
\item From $Y^{(s)}[0],\ldots,Y^{(s)}[t]$, the relay computes
$R[t]\in\mathbb F^k$ by its scheduled time. This may contain
original coordinates or extended innovations.
\item An independently valid $(b_2,T-b_1)$ encoder maps
$R[0],\ldots,R[t]$ to $Z[t]=(R[t],P'[t])\in\mathbb F^n$.
The destination uses the received RD packets to recover $S[t-T]$.
\end{itemize}
Both components have rate $k/n$. The following definition specifies
the non-adaptive benchmark used in this paper.

\begin{definition}[Fixed-rate non-adaptive TBSC]\label{def:nonadaptive}
Packet dimensions and the encoding schedule are fixed before
transmission. For every fixed source sequence, the relay transmits
the same $Z[t]$ at every time under every admissible SR erasure
history. It may use its observations to recover the required
values, but those prescribed output values do not depend on the
erasure history. Internal decoder state and recovery computations
may depend on that history; only transmitted output values
and the prescribed schedule must be history independent.
\end{definition}

In each construction below, every $R[t]$ has a fixed expression
in the source stream and is available at its scheduled time.
The fixed RD encoder then produces an erasure-state-independent
output. A missing input is never replaced by an erasure-dependent
placeholder.
\tikzset{
	node style/.style = {
		draw, circle, minimum size=1.5cm, thick,
		font=\large\bfseries,
	},
	label node/.style = {
		font=\normalsize,
		inner sep=2pt
	}
}
\begin{figure}[htbp]
\centering
\begin{tikzpicture}[>=stealth,every node/.style={font=\normalsize}]
\node[draw,circle,minimum size=1.1cm] (s) at (0,0) {$\mathsf s$};
\node[draw,circle,minimum size=1.1cm] (r) at (4,0) {$\mathsf r$};
\node[draw,circle,minimum size=1.1cm] (d) at (8,0) {$\mathsf d$};
\node[left=0.25cm of s] {$S[t]\in\mathbb F^k$};
\node[right=0.25cm of d] {$\widetilde S[t-T]\in\mathbb F^k$};
\draw[->] (s)--(r) node[midway,above=4pt] {$X[t]\in\mathbb F^n$};
\draw[->] (r)--(d) node[midway,above=4pt] {$Z[t]\in\mathbb F^n$};
\node[align=center] at (2,-0.9) {Burst erasure\\of at most $b_1$ packets};
\node[align=center] at (6,-0.9) {Burst erasure\\of at most $b_2$ packets};
\end{tikzpicture}
\caption{Three-node relay network with burst-erasure parameters $(b_1,b_2,T)$.}
\label{fig:relay}
\end{figure}
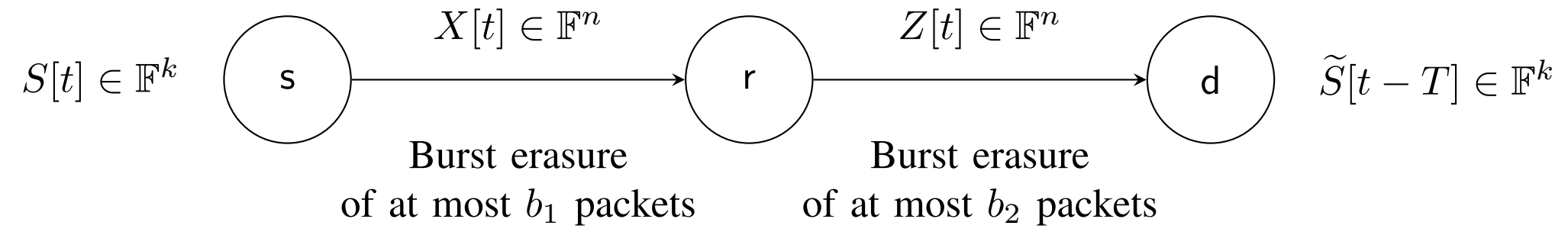

\begin{definition}\label{d2}
		Let $S[t]=(s_1[t],\ldots,s_k[t])$. Suppose that, for every
		admissible erasure pattern, every $t\geq0$, and every $i\in[k]$,
		the receiver recovers by $t+t_i$ an innovation
		\(
		s_i[t] + m_i(t),
		\)
where
\[
m_i(t)=\sum_{\ell=1}^{L}\sum_{j=1}^{k}
\alpha_{ij\ell}s_j[t-\ell].
\]
The finite memory $L$ and all coefficients $\alpha_{ij\ell}$ are
fixed before transmission, do not vary with $t$, and are independent
of the erasure history. Inputs at negative times are fixed to zero.
		
		The vector $(t_1,\ldots,t_k)$, together with the corresponding innovations
		\[
		\bigl(s_1[t]+m_1(t),\ldots,s_k[t]+m_k(t)\bigr),
		\]
is called an \emph{extended delay profile}. The $t_i$ are guaranteed
upper bounds; minimality is not part of a construction claim.
The transformation is causal and unitriangular in time, and is
therefore invertible on every finite time horizon.
	\end{definition}

\begin{remark}[Relation to ordinary delay profiles]
An ordinary delay profile is obtained by setting $m_i(t)=0$ for every $i$ and $t$. Hence ordinary profiles form the identity-transform subclass of Definition~\ref{d2}. The inclusion is strict at the representation level: for example, the innovation $s_1[t]+s_2[t-1]$ is allowed by Definition~\ref{d2} but is not an ordinary coordinate. The substantive role of this extension is not merely notational; in Section~\ref{ex:new-only}, it yields a rate-optimal code for $(b_1,b_2,T)=(3,4,12)$, a parameter triple outside the earlier large-delay region.
\end{remark}

We use three compositions: ordinary profiles on both links, an
extended profile on SR, or a separable extended profile on RD.
A fixed permutation of RD input coordinates is always allowed.
The ordinary composition states it explicitly; for the extended
compositions we relabel RD inputs first, so their displayed maps
use aligned indices without restricting the allowed matching.

\noindent \textit{Case 1: Ordinary profiles on both links.}

\begin{lemma} \label{case1}
If the SR and RD components have ordinary profiles
$(t_i)_{i\in[k]}$ and $(t'_j)_{j\in[k]}$, and a permutation
$\pi$ satisfies $t_i+t'_{\pi(i)}\leq T$ for every $i$,
then ordinary forwarding recovers every source packet within $T$.
\end{lemma}
\begin{proof}
At time $z$, the relay knows $s_i[z-t_i]$ and sends it in RD
input position $\pi(i)$:
$R_{\pi(i)}[z]=s_i[z-t_i]$.
The RD guarantee recovers that input by $z+t'_{\pi(i)}$.
Thus $s_i[\tau]$ is recovered by
$\tau+t_i+t'_{\pi(i)}\leq\tau+T$.
\end{proof}
\noindent \textit{Case 2: Extended SR and ordinary RD profiles.}

We next use an extended delay profile on the SR link and an ordinary profile on the RD link.

\begin{lemma}\label{case2}
Suppose that the SR link has an extended delay profile $(t_1,\ldots,t_k)$ for the innovations $s_i[t]+m_i(t)$ and that the RD link has an ordinary delay profile $(t'_1,\ldots,t'_k)$. If $t_i+t'_i\leq T$ for every $i\in[k]$, then every source symbol is recovered at the destination within delay $T$.
\end{lemma}

\begin{proof}
	At time $t$, the relay sets
	$r_i[t]=s_i[t-t_i]+m_i(t-t_i)$, which it has already recovered.
	The RD decoder obtains $r_i[t]$ by $t+t'_i\leq(t-t_i)+T$.
	Induct on the source time $\tau=t-t_i$. Every term in
	$m_i(\tau)$ has source time below $\tau$ and is known by its
	own deadline, no later than $\tau+T$. Subtracting it from
	$r_i[t]$ recovers $s_i[\tau]$ by $\tau+T$.
\end{proof}

\noindent \textit{Case 3: Ordinary SR and separable extended RD profiles.}

For an ordinary SR profile paired with an extended RD profile, we use
the following separability property.
\begin{definition}\label{def:eds}
	An extended delay profile for an input stream
	$W[t]=(w_1[t],\ldots,w_k[t])$ is \emph{separable} if $[k]$ can
	be partitioned as $A_I\mathbin{\dot\cup}A_D$ such that its
	innovations are $w_j[t]$ for $j\in A_I$ and $w_i[t]+m_i(t)$
	for $i\in A_D$, where $m_i(t)$ is a linear combination of
	$\{w_j[t-d]:d>0,\ j\in A_I\}$.
	
	For example, $(r_1[t]+r_3[t-1]+r_4[t-1],r_2[t]+r_3[t-2],r_3[t],r_4[t])$ is separable with $A_D=\{1,2\}$ and $A_I=\{3,4\}$.
\end{definition}
\begin{lemma} \label{case3}
Suppose that the SR link has an ordinary profile $(t_1,\ldots,t_k)$ and the RD link has a separable extended profile $(t'_1,\ldots,t'_k)$. If $t_i+t'_i\leq T$ for every $i\in[k]$, then all source symbols can be recovered at the destination within delay $T$.
\end{lemma}
\begin{proof}
	By time $t$, the relay knows $s_i[t-t_i]$ for every $i$.
	For $j\in A_I$, set $r_j[t]=s_j[t-t_j]$. All terms of
	$m_i(t)$, $i\in A_D$, are past values of these independent
	coordinates and are therefore available at the relay. Set
	$r_i[t]=s_i[t-t_i]-m_i(t)$ for $i\in A_D$. The RD decoder
	obtains $r_j[t]=s_j[t-t_j]$ for $j\in A_I$ and
	$r_i[t]+m_i(t)=s_i[t-t_i]$ for $i\in A_D$ by $t+t'_i$.
	Each deadline is at most $(t-t_i)+T$.
\end{proof}

\subsection{Rate Upper Bound}
The following bound is the burst-only specialization of \cite[Theorem~1]{9834645}. We include a short proof to identify the use of state-independent relaying.
\begin{theorem}\label{thm:rate-upper-bound}
	The rate $R$ of any fixed-rate non-adaptive $(b_1,b_2,T)$ TBSC satisfies
		\begin{align}  
			R &\leq  R(b_1,b_2,T) \notag \\ 
			&= \begin{cases}
				\min \biggl\{ \dfrac{T-b_1}{T-b_1+b_2}, \dfrac{T-b_2}{T-b_2+b_1} \biggr\} 
				& \text{if } T \geq b_1+b_2, \\
				0 & \text{otherwise}.
			\end{cases}
		\end{align}
	\end{theorem}
	\begin{proof}
	If $T<b_1+b_2$, erase SR packets at times $0,\ldots,b_1-1$
	and RD packets at times $b_1,\ldots,b_1+b_2-1$. No destination
	observation through time $T$ can depend on $S[0]$, so a positive
	rate is impossible.

	Now let $T\geq b_1+b_2$. Lemma~\ref{lemma:relay-decoding} and burst separation turn the SR encoder into a
	$(b_1,T-b_2)$ point-to-point code. Its burst converse gives
	$R\leq(T-b_2)/(T-b_2+b_1)$ without state-independent relaying.

	For the other term, fix a source sequence. By Definition~\ref{def:nonadaptive},
	$Z[\tau]$ is the same under every admissible SR history. If
	$\tau<b_1$, erase SR packets from $0$ through $\tau$; the relay
	has no source-dependent observation by $\tau$. If $\tau\geq b_1$,
	erase the $b_1$ SR packets from $\tau-b_1+1$ through $\tau$.
	By causality, the relay then has no observation depending on
	$S[j]$ with $j>\tau-b_1$. State independence therefore makes
	$Z[\tau]$ a function only of $S[0],\ldots,S[\tau-b_1]$
	for every SR history, with an empty input list when $\tau<b_1$.
	Consequently, $Z^{\star}[j]=Z[j+b_1]$ is a causal point-to-point
	encoder for the source stream. Prepending $b_1$ fixed, unerased
	RD packets to any isolated RD burst for $Z^{\star}$ reproduces
	an admissible original RD history. The original destination decoder
	then recovers $S[t]$ from this shifted stream by
$t+T-b_1$. The burst-separation argument in Section~\ref{sec:component-checks} makes this a
	$(b_2,T-b_1)$ point-to-point code, whose burst converse gives
	$R\leq(T-b_1)/(T-b_1+b_2)$. Combining the two inequalities proves
	the claim.
	\end{proof}
	When $b_1\geq b_2$, the SR inequality in the proof is the
	active term, so the same numerical bound holds even for an
	arbitrary causal, possibly state-dependent relay. For $b_1<b_2$,
	the RD inequality requires the state-independence condition in
	Definition~\ref{def:nonadaptive}.
	
	A non-adaptive $(b_1,b_2,T)$ TBSC is \emph{rate-optimal} if its rate equals $R(b_1,b_2,T)$.
	
\subsection{General Achievability Result}
The following theorem states the enlarged sufficient range.
The constructions below explain how the relay meets the deadlines.
\begin{theorem}[Binary attainment with SR-2]
	\label{pw:achievability}\label{thm:SR2-achievable}
	Let $0<v<u$, $p\ge2$, $0<q<u$, and $T-v=pu+q$. If
	\begin{equation}
		q(v-q)\le(p-1)u(u-v),\label{eq:main-tail-condition}
	\end{equation}
	a binary fixed-rate non-adaptive code of dimensions
	$(k,n)=(T-v,T-v+u)$ attains $R^\star=(T-v)/(T-v+u)$
	within delay $T$, for either link order $(v,u)$ or $(u,v)$.
\end{theorem}
\begin{proof}
	Lemma~\ref{pw:sr2-profile-lemma} gives the SR-2 extended profile;
	Lemma~\ref{RD} gives the common ordinary profile on the longer-burst
	hop. Their pairing and the relay maps are verified in
	Appendix~\ref{app:sr2-pairing} for both orders.
\end{proof}
The \arxivfix{SR range} $q\le A=(p-1)(u-v)$ is included because
$q(v-q)\le qu\le uA$ there. For example, $(v,u,T)=(5,6,20)$
has $p=2$, $q=3>A=1$, but satisfies \eqref{eq:main-tail-condition}
with equality. Section~\ref{ex:sr2-56} writes out that code.
The case $q=0$, including equal bursts, uses the ordinary code
in Theorem~\ref{thm:extended-achievability}.

\section{Binary Construction I: SR and the Common RD Code}\label{KEY CONSTRUCTION}
Write $T-v=pu+q$. In the nondivisible case assume
\begin{equation}
u>v,\quad p\ge2,\quad0<q<u.
\label{eq:original-parameter-range}
\end{equation}
We first describe SR on the shorter-burst hop, then the common RD
encoder on the longer-burst hop. Swapping the components uses
the separable composition of Lemma~\ref{case3}.
\srnamefix{SR and SR-2 name the two shorter-burst constructions.
The terms SR link and RD link identify the physical network hops.}
\subsection{The SR Component}
Set $(b_1,b_2)=(v,u)$, $H=pv+q$, and $\delta=(p-1)v$.
The source order is
\[
 S[t]=(V[t],B[t],U[t],(G_j[t])_{j=p-2,\ldots,1}),
 \qquad |V|=u,\quad |B|=q,\quad |U|=|G_j|=u.
\]
For $1\le s<v$, define a $q\times u$ matrix $\fourthfix{\mathsf{L}_s}$.
If $q<v$, its only nonzero entry is
\[
 \fourthfix{\mathsf{L}_s}(\langle s\rangle_q,q+\langle s\rangle_{v-q})=\nz1;
\]
if $q\ge v$, let $\fourthfix{\mathsf{L}_s}=0$. Put $J=(I_q\mid0)$.
The \arxivfix{encoding matrices}, in source-group order, are
\begin{equation}
 P_\ell=\begin{pmatrix}
 \ind{\ell=H}\nz{I_u}\\
 \ind{\ell=pv}\nz J+\displaystyle\sum_{s=1}^{v-1}
                            \ind{\ell=\delta+s}\nz{\fourthfix{\mathsf{L}_s}}\\
 \ind{\ell=\delta}\nz{I_u}\\
 (\ind{\ell=jv}\nz{I_u})_{j=p-2,\ldots,1}
 \end{pmatrix}_{(pu+q)\times u}.
 \label{eq:original-sr-taps}
\end{equation}
All unassigned entries are zero. The source sends
$P[t]=\sum_\ell S[t-\ell]P_\ell$ systematically.
Thus each $P_\ell$ has exactly $k=pu+q$ rows.
\thirdfix{In its $B$ block row, $J$ and every $\fourthfix{\mathsf{L}_s}$ are $q\times u$.
The same $\fourthfix{\mathsf{L}_s}$ appear in the correction $M_0[t]$ in
Lemma~\ref{delay SR}.}

Consider a burst at $0,\ldots,v-1$ and clean the received
parities. Retain parity times $v,\ldots,H+v-1$; later
columns have no erased-input contribution. Partition them into
the windows $[jv,(j+1)v-1]$, $1\le j\le p-2$,
$[\delta,pv-1]$, $[pv,H-1]$, and $[H,H+v-1]$.
Regrouping the unknowns as $(\mathbf G_1,\ldots,
\mathbf G_{p-2},\mathbf U,\mathbf B,\mathbf V)$ gives
\begin{equation}
 \widehat{\mathbf P}_{\rm SR}=
 \left(\begin{array}{c|c|cc}
 \nz{I_{(p-2)vu}}&0&0&0\\ \hline
 0&\nz{I_{vu}}&0&0\\ \hline
 0&\nz{B_U}&\nz{B_0}&\nz{B_1}\\
 0&0&0&\nz{I_{vu}}
 \end{array}\right).
 \label{eq:original-recovery-blocks}
\end{equation}
The three $B$ blocks simply place the assigned taps of the
$B$ row of \eqref{eq:original-sr-taps} in those three
consecutive windows. The column widths are
$(p-2)vu,vu,qu,vu$ and row heights are
$(p-2)vu,vu,vq,vu$.

\begin{figure}[!htbp]
\centering
\begin{tikzpicture}[every node/.style={font=\small},x=1.9cm,y=0.82cm]
\node at (0.5,0.45) {$\mathbf P_0,\ldots,\mathbf P_{p-3}$};
\node at (1.5,0.45) {$\mathbf P_{p-2}$};
\node at (2.5,0.45) {$\mathbf P_{p-1}$};
\node at (3.5,0.45) {$\mathbf P_p$};
\foreach \y/\lab in {0/{$\mathbf G_1,\ldots,\mathbf G_{p-2}$},1/{$\mathbf U$},2/{$\mathbf B$},3/{$\mathbf V$}} {
\node[anchor=east] at (-0.15,-\y-0.5) {\lab};}
\fill[assigned!15] (0,0) rectangle (1,-1);
\fill[assigned!15] (1,-1) rectangle (2,-2);
\fill[assigned!25] (1,-2) rectangle (4,-3);
\fill[assigned!15] (3,-3) rectangle (4,-4);
\draw (0,0) rectangle (4,-4);
\foreach \x in {1,2,3} {\draw (\x,0)--(\x,-4);}
\foreach \y in {1,2,3} {\draw (0,-\y)--(4,-\y);}
\node at (0.5,-0.5) {$I$}; \node at (1.5,-1.5) {$I$};
\node at (1.5,-2.5) {$B_U$};\node at (2.5,-2.5) {$B_0$};
\node at (3.5,-2.5) {$B_1$};\node at (3.5,-3.5) {$I$};
\node[anchor=west] at (4.15,-0.5) {$G_j$: direct};
\node[anchor=west] at (4.15,-1.5) {$U+M_0$: retain mixture};
\node[anchor=west] at (4.15,-2.5) {$B$: transition recovery};
\node[anchor=west] at (4.15,-3.5) {$V$: subtract $B$, then read};
\draw[-{Stealth}] (0,-4.5)--(4,-4.5) node[midway,below] {received parity windows in time order};
\end{tikzpicture}
\caption{Schematic of $\mathbf P_{\rm SR}$ after regrouping source
rows. Blank cells are zero; shaded cells contain the assigned blocks.
The first column group is absent for $p=2$. The time windows and
block dimensions are given above.
The diagram shows the recovery order without expanding every tap.}
\label{figure PSR}
\end{figure}
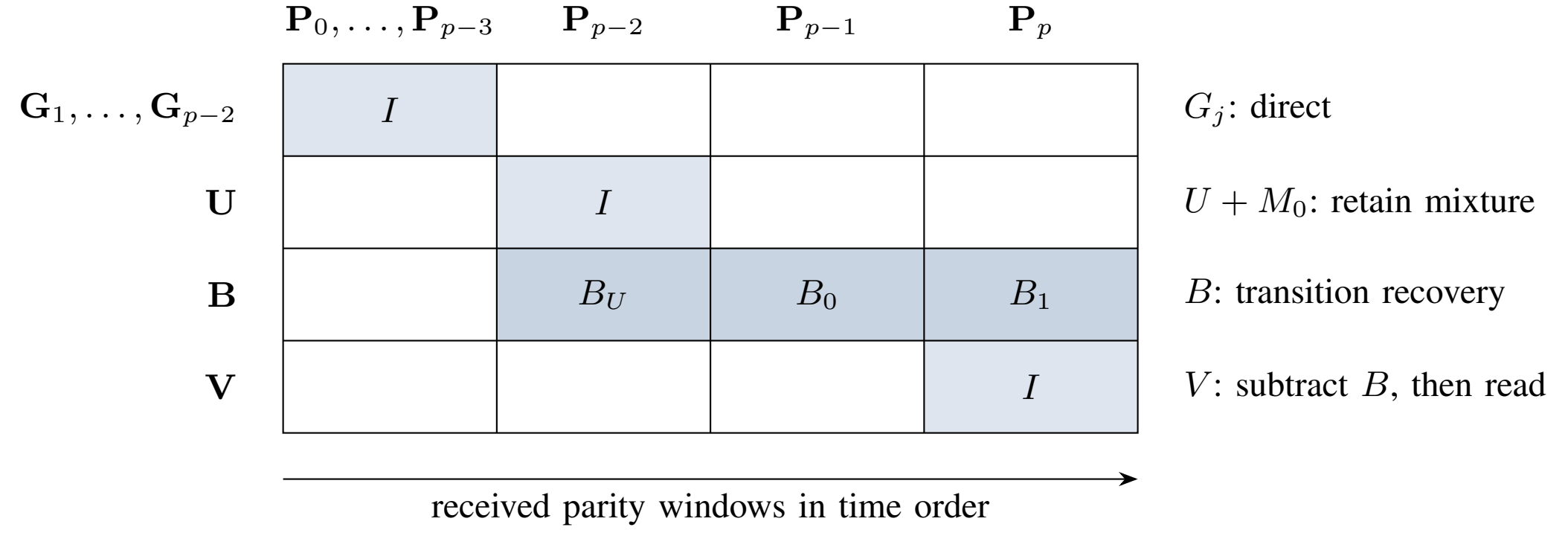

\begin{lemma}[The SR transition block]\label{lemma:P_invertible}
If $0<q<v$, the $vq\times vq$ matrix obtained from $B_0$
by retaining the first $v$ parity coordinates per time is
invertible. All erased $B$ packets are known by $H-1$, and
each $B[i]$ is known by $pv+i$. If $q\ge v$, the main
parities give $B[i]$ directly at $pv+i$ instead.
\end{lemma}
The transition-block proof is given in Appendix~\ref{app:sr-transition}.

\begin{samepage}
\begin{lemma}[The SR extended profile]\label{delay SR}
The code \eqref{eq:original-sr-taps} has the separable
extended profile
\[
 (H\mathbf1_u,pv\mathbf1_q,\delta\mathbf1_u,
                              (jv\mathbf1_u)_{j=p-2,\ldots,1}).
\]
The $V,B,G_j$ symbols are ordinary. For $U$ use the fixed
innovation
\begin{equation}
 I_U[t]=U[t]+M_0[t],\qquad
 M_0[t]=\sum_{s=1}^{v-1}B[t-s]\fourthfix{\mathsf{L}_s}.
 \label{eq:sr-explicit-innovation}
\end{equation}
In particular, $M_0=0$ when $q\ge v$.
\end{lemma}
\end{samepage}
\begin{proof}
We read \eqref{eq:original-recovery-blocks} in its column order.

\noindent\emph{1) The early identity blocks.}
The first blocks give $G_j[i]$ at $jv+i$. In the next window,
the parity at $\delta+i$ gives $U[i]+M_0[i]$, after
subtracting the known term $B[i-v]J$ and adding back known
outside-burst contributions to $M_0[i]$. Thus the erased
$U$ packets have their stated innovations by $\delta+i$.

\noindent\emph{2) The transition block.}
Lemma~\ref{lemma:P_invertible} recovers $B[i]$ by $pv+i$,
and recovers all erased $B$ packets before $H$.

\noindent\emph{3) The final identity block.}
At $H+i$, remove the now known $B$ contribution $B_1$.
The $V$ identity block gives $V[i]$ directly, proving delay
$H$. This ordinary recovery uses only received outside-burst
inputs and the recovered $B$ group; it does not assume that
an outside-burst innovation is already known.

\begin{refineblock}
\noindent\emph{4) The innovation is available at every source time.}
Put $z=t+\delta$. If the parity packet at $z$ is received,
its terms $V[z-H]$, $B[t-v]J$, and $G_j[z-jv]$ are known
by their ordinary deadlines, all at most $z$.
Subtracting them gives exactly $I_U[t]$.
If that packet is erased in a burst starting at $c$, then
$\delta\ge v$ implies $t=z-\delta<c$.
Thus $U[t]$ and every older $B$ symbol in $M_0[t]$ precede
this isolated burst and are known systematically.
The ordinary recoveries in Steps 1--3 are independent of this
check, so there is no circular dependency.
\end{refineblock}

Every correction term in \eqref{eq:sr-explicit-innovation}
is an older symbol of the ordinary $B$ group, proving
separability. All original inputs are known by delay $H$,
including $U$ after its older $B$ terms arrive. Time translation,
shorter-burst padding up to the relevant individual deadlines,
and the guard interval extend the proof to the stated channel.
\end{proof}

\subsection{The Common RD Code}
\label{sec:shared-rd}

Both constructions use the following ordinary code on the
longer-burst hop. \arxivfix{Its delay profile determines what the SR component must recover and when.}
Let its burst length be $B$, let $k=pB+q$, and first suppose
$0<q<B$. Split its input as
\[
 R[t]=(F[t],W_1[t],\ldots,W_p[t]),\qquad
 |F|=q,\quad |W_a|=B.
\]
For $1\le\ell<B$, the $q\times(B-q)$ matrix $V_\ell$ has
one assigned entry,
\begin{equation}
 V_\ell(\langle\ell\rangle_q,\langle\ell\rangle_{B-q})=\nz1,
 \qquad\langle\ell\rangle_s=1+((\ell-1)\bmod s),
 \label{eq:shared-rd-cyclic}
\end{equation}
and all its other entries are zero. Put $V_\ell=0$ otherwise.
The encoding matrices are
\begin{equation}
 P'_\ell=
 \left(\begin{array}{c@{\quad}c}
 \ind{\ell=B}\nz{I_q}&\nz{V_\ell}\\ \hline
 \multicolumn{2}{c}{(\ind{\ell=aB+q}\nz{I_B})_{a=1,\ldots,p}}
 \end{array}\right)_{k\times B},\qquad
 P'_\ell=0\quad(\ell\notin[0,k]).
 \label{eq:shared-rd-taps}
\end{equation}
The transmitted parity is $Q[t]=\sum_\ell R[t-\ell]P'_\ell$.
For a burst at $0,\ldots,B-1$, subtract the known outside-burst
inputs from the received parities. Their recovery matrix is
\begin{equation}
 \begin{gathered}
 (\bar Q[B],\ldots,\bar Q[k+B-1])
                =(R[0],\ldots,R[B-1])\mathbf P_{\rm RD},\\
 \mathbf P_{\rm RD}=(P'_{B+s-i})_{0\le i<B,\,0\le s<k}.
 \end{gathered}
 \label{eqRDP}
\end{equation}
Partition its parity times into the head window
$\mathcal W_0=[B,B+q-1]$ and the $B$-packet windows
$\mathcal W_a=[aB+q,(a+1)B+q-1]$, $1\le a\le p$.
Regrouping the erased input coordinates by $F,W_1,\ldots,W_p$
gives the block matrix
\begin{equation}
 \widehat{\mathbf P}_{\rm RD}=
 \begin{array}{c|ccccc}
 &\mathcal W_0&\mathcal W_1&\mathcal W_2&\cdots&\mathcal W_p\\ \hline
 \mathbf F&\nz{H_0}&\nz{H_1}&0&\cdots&0\\
 \mathbf W_1&0&\nz{I_{B^2}}&0&\cdots&0\\
 \mathbf W_2&0&0&\nz{I_{B^2}}&\cdots&0\\
 \vdots&\vdots&\vdots&\vdots&\ddots&\vdots\\
 \mathbf W_p&0&0&0&\cdots&\nz{I_{B^2}}
 \end{array}.
 \label{eq:shared-rd-blocks}
\end{equation}
Here bold groups concatenate their $B$ erased input packets.
With $K_\ell=(\ind{\ell=B}I_q\mid V_\ell)$, every entry of
the two remaining blocks is specified by
\[
 H_0=(K_{B+s-i})_{0\le i<B,\,0\le s<q},\qquad
 H_1=(K_{B+q+s-i})_{0\le i<B,\,0\le s<B}.
\]
In particular, $H_0$ is $Bq\times Bq$.

\begin{lemma}[Recovery of the RD head]\label{invertible}
For $0<q<B$, the head block $H_0$ is invertible over
$\mathbb F_2$. It recovers $F[i]$ by $B+i$ for every
$0\le i<B$.
\end{lemma}
\begin{proof}
For $i<q$, the first $q$ parity coordinates at $B+i$ give
$F[i]$ directly. By $B+q-1$ all these prefix packets are known.
Subtract their contributions from the last $B-q$ coordinates
of the received head window. The remaining matrix is a permutation
matrix: for a fixed received time $t\in[B,B+q-1]$, the remaining
input times $i=q,\ldots,B-1$ give $B-q$ consecutive lags $t-i$,
which occupy the $B-q$ output positions once each in
\eqref{eq:shared-rd-cyclic}. For a fixed input time $i$, the
$q$ received times give $q$ consecutive lags and occupy each
of its $q$ coordinates once. Thus every remaining bit is read
at one distinct position. These packets are known by
$B+q-1\le B+i$ for $i\ge q$. This proves invertibility together with each packet's deadline.
\end{proof}

\begin{lemma}[The ordinary RD profile]\label{RD}
The code \eqref{eq:shared-rd-taps} has ordinary delays
\begin{equation}
 \begin{array}{c|ccccc}
 \text{group}&F\ (q)&W_1\ (B)&W_2\ (B)&\cdots&W_p\ (B)\\ \hline
 \text{delay}&B&B+q&2B+q&\cdots&pB+q
 \end{array}.
 \label{eq:shared-rd-profile}
\end{equation}
For $q=0$, omit $F$ and all $V_\ell$ and retain the taps
$aB$, $1\le a\le p$; the delays are $B,2B,\ldots,pB$.
\end{lemma}
\begin{proof}
Lemma~\ref{invertible} recovers $F$ by the end of the head
window. Remove its contribution $H_1$ from the next window.
The identity block in $\mathcal W_a$ then gives $W_a[i]$
at time $aB+q+i$. There is no interference from another
$W$ group. When $q=0$ these identity blocks are the whole
matrix, so the same conclusion is immediate.
The ordinary guarantee also holds outside the erased interval
because those input packets are received systematically.
Time translation and burst separation give the sliding-window
guarantee for every input stream.
\end{proof}
\subsection{Pairing the Profiles}
Split $V=(F,A^{\rm o})$ with sizes $q,u-q$. In the shared RD
code set $B=u$, $W_1=(A^{\rm o},B)$, $W_2=U$, and
$W_a=G_{p-a+1}$ for $3\le a\le p$. No RD coefficient
changes under this grouping.
\begin{lemma}\label{Construction-2}
Under \eqref{eq:original-parameter-range}, these two components
attain the rate upper bound in either link order.
\end{lemma}
\begin{proof}
Their coordinatewise delay sums are
\[
\begin{array}{c|c|c|c}
	\text{group}&\text{short-hop delay}&\text{long-hop delay}&\text{sum}\\ \hline
	F&pv+q&u&pv+q+u\le T\\
	A^{\rm o}&pv+q&u+q&pv+q+u+q\le T\\
	B&pv&u+q&pv+u+q\le T\\
	U&(p-1)v&2u+q&(p-1)v+2u+q\le T\\
	G_j&jv&(p-j+1)u+q&jv+(p-j+1)u+q\le T
\end{array}.
\]
The $A^{\rm o}$ inequality is exactly
$q\le(p-1)(u-v)$; all the others follow from $u\ge v$.
For order $(v,u)$, forward the short-hop innovations at their
guaranteed times and use Lemma~\ref{case2}. For order $(u,v)$,
use the same components in reverse and precompensate their
separable correction as in Lemma~\ref{case3}. Both components
are valid for every input stream and have the same dimensions,
so both compositions apply. The rate is
$(pu+q)/((p+1)u+q)=R(b_1,b_2,T)$.
\end{proof}

\begin{thirdblock}
The ordinary RD encoder in Lemma~\ref{RD} is reused from
\cite{li2025rateoptimalstreamingcodesthreenode}, Section III-B.
The SR component in Section III-A couples $B$ to the parity
through $\fourthfix{\mathsf{L}_s}$ and recovers $U[t]+M_0[t]$ at delay $\delta$.
Lemmas~\ref{lemma:P_invertible} and~\ref{delay SR} establish its
transition block and extended profile. The following theorem
pairs that profile with the common ordinary RD component.
\end{thirdblock}
\begin{theorem}[The SR construction and divisible cases]
\label{thm:extended-achievability}
Let $T\ge u+v$ and write $T-v=pu+q$, $0\le q<u$.
If $q=0$, or if \eqref{eq:original-parameter-range} holds,
a binary fixed-rate non-adaptive code attains $R^\star$ in either
link order. Equivalently, the sufficient condition is
\[
\frac{T-u-v}{2u-v}\le\left\lfloor\frac{T-u-v}{u}\right\rfloor.
\]
For $u=v=b$, this reduces to $b\mid(T-2b)$.
\end{theorem}
\begin{proof}
For $q>0$, apply Lemma~\ref{Construction-2}.
For $q=0$, use $p$ groups of $u$ coordinates and identity taps
only. Pair the short-hop delays $(pv,(p-1)v,\ldots,v)$
with the long-hop delays $(u,2u,\ldots,pu)$.
Each sum $(p-a+1)v+au$ is at most $v+pu=T$.
Lemma~\ref{case1} gives either link order, including $u=v$.
Both components have $(k,n)=(pu,(p+1)u)$, proving the rate.
\refine{For the equivalent condition, note that
$\lfloor(T-u-v)/u\rfloor=p-1$; the displayed inequality
therefore reduces to $q\le(p-1)(u-v)$, including $q=0$.}
\end{proof}

\section{Binary Construction II: The SR-2 Delay Profile}
\label{sec:SR2construction}
\subsection{The SR-2 Component}
SR-2 uses the same dimensions and RD encoder as SR, but divides
the first $u+q$ source coordinates into three groups. Under the hypotheses of Theorem~\ref{pw:achievability}, set
\begin{equation}
\begin{gathered}
m=u-q,\quad A=(p-1)(u-v),\quad\delta=(p-1)v,\\
E=T-u-q=pv+A,\qquad D=T-u=E+q.
\end{gathered}\label{pw:derived}
\end{equation}
Here $D$ is the full-packet deadline of the short-burst component.
Use dimensions $(k,n)=(pu+q,pu+q+u)$.
Write the source packet in the order
\begin{thirdblock}
\begin{equation}
\begin{gathered}
S[t]=\bigl(F[t],A^{\rm o}[t],B[t],U[t],
                    (G_j[t])_{j=p-2,\ldots,1}\bigr),\\
|F|=|B|=q,\qquad |A^{\rm o}|=m,\qquad |U|=|G_j|=u,\\
k=q+m+q+u+(p-2)u=pu+q.
\end{gathered}
\label{eq:sr2-groups}
\end{equation}
\end{thirdblock}
The $G_j$ groups are absent for $p=2$.

Put $L=A+q=D-pv$ and $h=v-L$.
Let $J_F=(I_q\mid0_{q\times m})$ and
$J_A=(0_{m\times q}\mid I_m)$.
Choose the binary matrices $C_j\in\mathbb F_2^{m\times q}$ from
$C(z)=\sum_{j=0}^d C_jz^j$ by
\begin{equation}
 \fitmatrix{(d,\theta,C(z))=
 \begin{cases}
 (0,0,0_{m\times q}),&h\le0,\\[2pt]
 \left(0,h,\begin{pmatrix}\nz{I_q}\\0_{(m-q)\times q}\end{pmatrix}\right),
                           &h>0,\ m\ge q,\ A\ge h,\\[5pt]
 (h-A,h,\nz{\ZZ(q,m-q,h-A)^{\mathsf T}}),&h>A,\\[2pt]
 (A-h,A,\nz{\ZZ(m,q-m,A-h)}),&h>0,\ m<q.
 \end{cases}}
 \label{pw:coefficient-choice}
\end{equation}
Here $\ZZ$ is the binary polynomial zigzag matrix specified in
Appendix~\ref{pw:zigzag}. The hypothesis $qh\le mA$ ensures
$m>q$ in the third case and $A>h$ in the fourth, so all arguments
of $\ZZ$ are positive.
Set $\lambda_j=\delta+\theta-j$.
Every nonzero auxiliary coefficient has
$\delta<\lambda_j<pv$.
In source-group order, assign the encoding matrices by
\begin{equation}
\begin{gathered}
P_\ell=\begin{pmatrix}
\ind{\ell=D}\nz{J_F}\\[3pt]
\ind{\ell=E}\nz{J_A}\\[3pt]
\ind{\ell=pv}\nz{J_F}
 +\displaystyle\sum_{j=0}^{d}\ind{\ell=\lambda_j}\nz{C_j^{\mathsf T}}J_A\\[3pt]
\ind{\ell=\delta}\nz{I_u}\\[3pt]
\bigl(\ind{\ell=jv}\nz{I_u}\bigr)_{j=p-2,\ldots,1}
\end{pmatrix}\in\mathbb F_2^{k\times u},\\
P_\ell=0\qquad(\ell\notin[0,D]).
\end{gathered}
\label{pw:short-taps}
\end{equation}
The block-row heights are $q,m,q,u,(p-2)u$, and every block
has $u$ columns. The last block row stacks $G_{p-2},\ldots,G_1$
and is empty when $p=2$. All unassigned entries are zero.
The source uses the systematic encoder~\eqref{eq-parity} with
these fixed coefficients.

For a burst at $0,\ldots,v-1$, substitute these matrices into
the recovery framework~\eqref{eq-recovery}--\eqref{eq-matrixP}
with burst length $v$ and component deadline $D$. Denote the
resulting recovery matrix by $P_{\rm SR\text{-}2}$:
\begin{equation}
 \begin{gathered}
 (\bar P[v],\ldots,\bar P[D+v-1])
       =(S[0],\ldots,S[v-1])P_{\srnamefix{\rm SR\text{-}2}},\\
 P_{\srnamefix{\rm SR\text{-}2}}=(P_{v+s-i})_{\substack{0\le i<v\\0\le s<D}}
              \in\mathbb F_2^{vk\times Du},
 \end{gathered}
 \label{pw:sr2-matrix}
\end{equation}
Its block columns follow the received parity packets in time order.
Partition them into the consecutive windows
\begin{equation}
 \underbrace{[jv,(j+1)v-1]}_{j=1,\ldots,p-2},\quad
 [\delta,pv-1],\quad[pv,E-1],\quad[E,D-1],\quad[D,D+v-1].
 \label{pw:windows}
\end{equation}
For $1\le j\le p-1$, let $\mathbf P_{j-1}$ collect parity times
$jv,\ldots,(j+1)v-1$. Let $\mathbf P_{p-1}$ collect
$pv,\ldots,D+v-1$, split into windows of lengths $A,q,v$.
Write $\mathbf X=(X[0],\ldots,X[v-1])$ for a source group.
Regroup the erased rows in the order
$((\mathbf G_j)_{j=1}^{p-2},\mathbf U,\mathbf B,
\mathbf A^{\rm o},\mathbf F)$.
The resulting matrix has the block structure
\begin{equation}
 \widehat P_{\srnamefix{\rm SR\text{-}2}}=
 \left(\begin{array}{c|c|ccc}
 \nz{I_{(p-2)vu}}&0&0&0&0\\ \hline
 0&\nz{I_{vu}}&0&0&0\\ \hline
 0&\nz{\mathcal B_U}&\nz{\mathcal B_C}&\nz{\mathcal B_J}&\nz{\mathcal B_F}\\
 0&0&0&\nz{\mathcal A_J}&\nz{\mathcal A_F}\\
 0&0&0&0&\nz{\mathcal F_F}
 \end{array}\right).
 \label{pw:sr2-blocks}
\end{equation}
The row heights are $((p-2)vu,vu,vq,vm,vq)$ and column widths
are $((p-2)vu,vu,Au,qu,vu)$, totalling $vk$ rows and $Du$ columns.
The first row and column groups are absent for $p=2$.
Figure~\ref{fig:sr2-recovery-windows} displays this structure.
The $B$ blocks place the assigned $B$ taps in their corresponding
windows. Appendix~\ref{app:sr2} gives their individual entries.

\begin{inlinefigure}
\centering
\fitgraphic{\linewidth}{%
\begin{tikzpicture}[every node/.style={font=\small},x=1cm,y=0.82cm]
\node at (1.1,1.1) {$\mathbf P_0,\ldots,\mathbf P_{p-3}$};
\node at (3.0,1.1) {$\mathbf P_{p-2}$};
\node at (6.2,1.1) {$\mathbf P_{p-1}$};
\draw (3.85,0.72)--(3.85,0.82)--(8.55,0.82)--(8.55,0.72);
\node[font=\scriptsize] at (1.1,0.35) {$[jv,(j+1)v-1]$};
\node[font=\scriptsize] at (3.0,0.35) {$[\delta,pv-1]$};
\node[font=\scriptsize] at (4.6,0.35) {$[pv,E-1]$};
\node[font=\scriptsize] at (6.2,0.35) {$[E,D-1]$};
\node[font=\scriptsize] at (7.8,0.35) {$[D,D+v-1]$};
\foreach \y/\lab in {0/{$\mathbf G_1,\ldots,\mathbf G_{p-2}$},1/{$\mathbf U$},2/{$\mathbf B$},3/{$\mathbf A^{\rm o}$},4/{$\mathbf F$}} {
\node[anchor=east] at (-0.15,-\y-0.5) {\lab};}
\fill[assigned!15] (0,0) rectangle (2.2,-1);
\fill[assigned!15] (2.2,-1) rectangle (3.8,-2);
\fill[assigned!25] (2.2,-2) rectangle (3.8,-3);
\fill[assigned!25] (3.8,-2) rectangle (5.4,-3);
\fill[assigned!25] (5.4,-2) rectangle (7,-3);
\fill[assigned!25] (7,-2) rectangle (8.6,-3);
\fill[assigned!15] (5.4,-3) rectangle (7,-4);
\fill[assigned!15] (7,-3) rectangle (8.6,-4);
\fill[assigned!15] (7,-4) rectangle (8.6,-5);
\draw (0,0) rectangle (8.6,-5);
\foreach \x in {2.2,3.8,5.4,7.0} {\draw (\x,0)--(\x,-5);}
\foreach \y in {1,2,3,4} {\draw (0,-\y)--(8.6,-\y);}
\node at (1.1,-0.5) {$I$};
\node at (3,-1.5) {$I$};
\node at (3,-2.5) {$\mathcal B_U$};
\node at (4.6,-2.5) {$\mathcal B_C$};
\node at (6.2,-2.5) {$\mathcal B_J$};
\node at (7.8,-2.5) {$\mathcal B_F$};
\node at (6.2,-3.5) {$\mathcal A_J$};
\node at (7.8,-3.5) {$\mathcal A_F$};
\node at (7.8,-4.5) {$\mathcal F_F$};
\node[anchor=west] at (8.85,-0.5) {$G_j$ recovered directly};
\node[anchor=west] at (8.85,-1.5) {$U+M$ retained};
\node[anchor=west] at (8.85,-2.5) {$B$ prefix and tail recovery};
\node[anchor=west] at (8.85,-3.5) {$A^{\rm o}$ after subtracting $B$};
\node[anchor=west] at (8.85,-4.5) {$F$ after subtracting $B$};
\draw[-{Stealth}] (0,-5.6)--(8.6,-5.6)
node[midway,below=3pt] {received parity windows in time order};
\end{tikzpicture}}
\caption{\srnamefix{Schematic of $\widehat P_{\rm SR\text{-}2}$ after regrouping
the erased source rows. Each bold row group collects its $v$ erased packets.
The first row and column collect $1\le j\le p-2$ and are absent for $p=2$.
The last three column groups form $\mathbf P_{p-1}$.
Blank cells are zero, and shaded cells show the assigned blocks of
\eqref{pw:sr2-blocks}. Column widths are schematic.}}
\label{fig:sr2-recovery-windows}
\end{inlinefigure}

The auxiliary coefficients recover the $B$ tail left by the
direct identity observations. For $(b_1,b_2,T)=(5,6,20)$,
the delay-$5$ innovation mixes $U$ with older $B$ symbols.
For a burst at $0,\ldots,4$, three extra parity observations
at time $10$ recover the last missing packet $B[4]$.
Section~\ref{ex:sr2-56} gives the full timeline.

\subsection{The SR-2 Extended Profile}
\begin{revisionblock}
For a burst of length $v$, the first $q$ parity positions recover
a $B$ prefix of length $\min\{A+q,v\}$. If $h>0$, the tail has $qh$ unknown symbols, and the
last $m$ parity positions at times $pv,\ldots,E-1$ give $mA$
auxiliary observations. The hypothesis of
Theorem~\ref{pw:achievability} therefore reads
\[
 qh\le mA
 \quad\Longleftrightarrow\quad q(v-q)\le uA.
\]
The zigzag coefficients give full row rank to this finite system.
Lemma~\ref{lem:sr2-final-block} checks that recovery, including
any prefix subtraction, meets each individual deadline.
Appendix~\ref{ex:zigzag-new-range} works out fifteen tail symbols
from two received packets. If $h\le0$, prefix recovery suffices.
\end{revisionblock}

\begin{revisionblock}
The count $qh\le mA$ alone does not ensure recovery from a
causal finite window. The coefficients must make the actual
observed matrix have full row rank. Use the coefficients in
\eqref{pw:coefficient-choice}, with $C_j=0$ outside $[0,d]$.
For $h>0$, after subtracting the known
$B$ prefix, the last $m$ parity positions in the $A$ packets
at times $pv,\ldots,E-1$ give
\[
\bar{\mathbf y}_{\rm aux}=\mathbf b_{\rm tail}\mathcal M,
\qquad
\mathcal M=(C_{L+i-s-(v-\theta)}^{\mathsf T})_
                   {0\le i<h,\,0\le s<A}.
\]
Thus $\mathcal M$ has exactly $qh$ rows and $mA$ columns.
\reviewfix{With the prefix fixed, the $qh$ tail symbols are arbitrary.
Recovery from this specified auxiliary window is therefore possible
exactly when $\mathcal M$ has full row rank, which requires $qh\le mA$.
Lemma~\ref{pw:tail-rank} supplies coefficients meeting this requirement.
The subsequent deadline check and Lemma~\ref{pw:sr2-profile-lemma}
turn that window recovery into a valid extended profile.}
\fifthfix{Table~\ref{tab:sr2-cases} collects the coefficient choices
and their tail-recovery times.}
\end{revisionblock}
\begin{table}[!htbp]
\begin{thirdblock}
\centering\small
\caption{SR-2 coefficient cases under $qh\le mA$}
\label{tab:sr2-cases}
\begin{tabular}{|c|c|c|l|}\hline
Condition&$C(z)$&$\theta$&Tail recovery\\\hline
$h\le0$&$0_{m\times q}$&$0$&Prefix suffices\\\hline
$m\ge q$, $A\ge h>0$&$\begin{pmatrix}I_q\\0_{(m-q)\times q}\end{pmatrix}$&$h$&Direct, by $E-1$\\\hline
$h>A$&$\ZZ(q,m-q,h-A)^{\mathsf T}$&$h$&Pure window, by $E-1$\\\hline
$m<q$, $h>0$&$\ZZ(m,q-m,A-h)$&$A$&Prefix subtraction, by $D-1$\\\hline
\end{tabular}
\par\smallskip
\begin{minipage}{0.97\linewidth}\footnotesize
Here $L=A+q$, $h=v-L$, and $\ZZ$ is specified in
Appendix~\ref{pw:zigzag}. The inequality $qh\le mA$ forces
$m>q$ in the third row and $A>h$ in the fourth row.
Thus every call to $\ZZ(r,c,d)$ has $r,c,d>0$.
In particular, $m=q$ uses the constant case and never calls
$\ZZ$ with $c=0$. Equation~\eqref{pw:coefficient-choice}
gives the coefficient degree in each case.
\end{minipage}
\end{thirdblock}
\end{table}

\begin{revisionblock}
\begin{lemma}[\reviewfix{Full rank of the finite tail window}]
\label{pw:tail-rank}
\refine{If $qh\le mA$, the choices in
\eqref{pw:coefficient-choice} give}
$\operatorname{rank}\mathcal M=qh$ when $h>0$.
For $m\ge q$ the prefix term is zero and the tail is known by
$E-1$. For $m<q$ it is known by $D-1$ after prefix subtraction.
If $h\le0$, no auxiliary coefficients are needed.
\end{lemma}
\begin{proof}[Proof outline]
If $m\ge q$ and $A\ge h>0$, use a constant coefficient
matrix with a copy of $I_q$. The tail matrix contains $h$
independent copies of this identity.
The other two cases use the binary zigzag matrix $\ZZ$ in
Appendix~\ref{pw:zigzag}. Reverse the tail into a vector
polynomial $X_*(z)$ of degree less than $h$.

If $m<q$, put $d=A-h$ and
$C=\ZZ(m,q-m,d)$. The $A$ observations contain every coefficient
of $CX_*$, whose degree is at most $A-1$.
Every nonzero polynomial vector in the kernel of $C$ has degree
at least $\lfloor m(A-h)/(q-m)\rfloor\ge h$, by $mA\ge qh$.
Zero observations therefore force $X_*=0$.

If $h>A$, put $d=h-A$ and
$C=\ZZ(q,m-q,d)^{\mathsf T}$. Only product coefficients of
degrees $d,\ldots,h-1$ are observed. If they vanish, write
$CX_*=L_0+z^hH_0$, where both end polynomials have degree below $d$.
The dual matrix $K$ of Lemma~\ref{pw:kernel-lemma} satisfies
$KC=0$, has row degrees at most $A$, and has no nonzero
polynomial kernel vector of degree below $d$.
In $KL_0+z^hKH_0=0$, the first term has degree below $h$
and the second starts at degree $h$. They vanish separately,
so $L_0=H_0=0$. Full column rank of $C$ then gives $X_*=0$.
Appendix~\ref{pw:zigzag} proves these degree properties and
the stated prefix-subtraction times in full.
\end{proof}

\noindent\emph{Why these recovery times meet every deadline.}
The prefix packet $B[i]$ is read at $pv+i\le E+i$.
In the prefix-subtracted case the tail is known only at $D-1$,
but its source indices satisfy $i\ge L=A+q$, so
$D-1=E+q-1\le E+i$. In the other tail cases it is known by $E-1$.
The $A^{\rm o}$ identity starts at $E$: any interfering erased
$B$ term is either already in the recovered tail or, in the
prefix-subtracted case, has an index outside the burst and is
known systematically. Finally, the first $q$ positions at $D+i$
give $F[i]+B[L+i]$; the second term is known by $D-1$ or is
outside the burst. This gives ordinary delays $E,E,D$ for
$B,A^{\rm o},F$, independently of the early $U$ innovation.
Lemma~\ref{lem:sr2-final-block} verifies the interference indices,
and Appendix~\ref{app:sr2-profile-proof} checks innovations at
unerased source times and successive bursts.
\end{revisionblock}

\begin{lemma}[The SR-2 extended delay profile]
\label{pw:sr2-profile-lemma}
The SR-2 code has the separable extended profile
\begin{equation}
\bigl(\underbrace{D,\ldots,D}_q,
\underbrace{E,\ldots,E}_m,\underbrace{E,\ldots,E}_q,
\underbrace{\delta,\ldots,\delta}_u,
(\underbrace{jv,\ldots,jv}_u)_{j=p-2,\ldots,1}\bigr).
\label{pw:short-profile}
\end{equation}
The groups $F,A^{\rm o},B,G_j$ are recovered individually.
The $U$ group is recovered as
\begin{equation}
I_U[t]=U[t]+M[t],\quad
M[t]=B[t-v]J_F+
\sum_{j=0}^d B[t-(\theta-j)]C_j^{\mathsf T}J_A.
\label{pw:innovation}
\end{equation}
Every term of $M[t]$ has a strictly earlier source time.
All original source coordinates are also recovered within delay $D$.
\end{lemma}
\begin{proof}
Read the recovery matrix in column order. The early identity blocks
give $G_j$ and the $U$ innovation. The final block recovers the
$B$ prefix and tail, then $A^{\rm o}$ and $F$ at delays $E,E,D$.
Lemma~\ref{lem:sr2-final-block} proves this final-block recovery;
Appendix~\ref{app:sr2-profile-proof} completes the blockwise argument
and checks innovations at source times outside the burst.
\end{proof}
Apply Lemma~\ref{RD} with burst parameter $u$ and identify
$W_1=(A^{\rm o},B)$,
$W_2=U$, and $W_a=G_{p-a+1}$. The corresponding long-hop delays
are $u,u+q,u+q,2u+q,(p-j+1)u+q$.
Each paired sum is at most $T$; Appendix~\ref{app:sr2-pairing}
gives the fixed relay maps. Section~\ref{ex:sr2-56} illustrates
the profile without requiring the general matrix construction.

\subsection{An Identity-Coefficient Example: $(5,6,20)$}\label{ex:sr2-56}
Take $(b_1,b_2,T)=(5,6,20)$ and $(k,n)=(15,21)$, so
$R^\star=5/7$. Write
$S[t]=(F[t],A^{\rm o}[t],B[t],U_F[t],U_A[t])$ with three
coordinates in each group, and put $U=(U_F,U_A)$.

The parameters used below are
\[
(v,u,p,q)=(5,6,2,3),\quad(A,m,L,h)=(1,3,4,1),
\qquad(\delta,E,D)=(5,11,14).
\]
Here $L=4$ is the directly recovered $B$ prefix length, $h=1$
is the tail length, and $mA=3$ extra observations recover its
$qh=3$ symbols. The SR condition fails since $q=3>A=1$.
The layered baseline condition also fails since
$T-u=14<5\lceil15/6\rceil=15$. All operations are binary.

\subsubsection{The SR-2 encoder and recovery}
The identity-coefficient encoder sends the six parity symbols
\begin{equation}
\begin{split}
P_F[t]&=F[t-14]+B[t-10]+U_F[t-5],\\
P_A[t]&=A^{\rm o}[t-11]+B[t-6]+U_A[t-5].
\end{split}\label{eq:sr2-56-parity}
\end{equation}
For a burst at $0,\ldots,4$, remove the known systematic
contributions. The remaining equations are read as follows.
\begin{center}
\begin{tabular}{c|l|l}
Received time&Cleaned parity&Recovered quantity\\\hline
$5+i$, $0\le i<5$&the $U$ identity and older $B$ terms&$I_U[i]$\\
$10,11,12,13$&$\bar P_F[10+i]=B[i]$&$B[0],\ldots,B[3]$\\
$10$&$\bar P_A[10]=B[4]$&the remaining $B$ packet\\
$11+i$, $0\le i<5$&$\bar P_A[11+i]=A^{\rm o}[i]$&$A^{\rm o}[i]$\\
$14+i$, $0\le i<5$&$\bar P_F[14+i]=F[i]+B[4+i]$&$F[i]$
\end{tabular}
\end{center}
At $11+i$, the possible $B[5+i]$ term is outside the burst.
At $14+i$, $B[4]$ is already known at time $10$ and the
other $B[4+i]$ terms are systematic. Thus every subtraction
uses an available value. The fixed early innovation is
\[
I_U[t]=\bigl(U_F[t]+B[t-5],\ U_A[t]+B[t-1]\bigr).
\]
Its guarantee also holds at unerased source times by the all-time
check in Appendix~\ref{app:sr2-profile-proof}.

\subsubsection{The long-burst component and end-to-end deadline}
Order the RD inputs as $R=(R_F,R_A,R_B,R_{U_F},R_{U_A})$.
Its $3+6+6$ grouping is precisely $F,W_1,W_2$ in
Lemma~\ref{RD}. Put $R_U=(R_{U_F},R_{U_A})$.
For $1\le a\le3$, the six parity coordinates are
\begin{align*}
Q_a[t]&=R_{F,a}[t-6]+R_{A,a}[t-9]+R_{U_F,a}[t-15],\\
Q_{3+a}[t]&=\sum_{\substack{1\le\ell\le5\\\langle\ell\rangle_3=a}}
 R_{F,\langle\ell\rangle_3}[t-\ell]
 +R_{B,a}[t-9]+R_{U_A,a}[t-15].
\end{align*}
The relay uses the fixed map
\[
R_F[t]=F[t-14],\quad R_A[t]=A^{\rm o}[t-11],\quad
R_B[t]=B[t-11],\quad R_U[t]=I_U[t-5].
\]
The paired deadlines are
\[
\begin{array}{c|rrrr}
\text{group}&F&A^{\rm o}&B&U\\\hline
\text{SR-2 delay}&14&11&11&5\ \text{(extended)}\\
\text{RD delay}&6&9&9&15\\
\text{sum}&20&20&20&20
\end{array}.
\]
By $t+20$, the destination has $I_U[t]$. It has already
recovered $B[t-5]$ by $t+15$ and $B[t-1]$ by $t+19$,
so it subtracts them to recover $U[t]$ within the same deadline.
For order $(6,5)$, the relay adds the corresponding older $B$
terms before the short-burst encoder. They cancel its innovations
as in Appendix~\ref{app:sr2-pairing}, giving the same deadline.

\subsection{Two Packets Recover Fifteen Tail Symbols}
\label{sec:zigzag-main}
For $(v,u,T)=(10,12,37)$, use groups $(F,A^{\rm o},B,U)$
of sizes $(3,9,3,12)$ and packet dimensions $(k,n)=(27,39)$.
The main parity recovers $B[0],\ldots,B[4]$, leaving fifteen
tail symbols in $B[5],\ldots,B[9]$.
The last nine parity positions at times $20,21$ give eighteen
observations. Write their cleaned vectors as $y_0,y_1$ and
$B=(B_1,B_2,B_3)$.

With the binary coefficient assignment in
Appendix~\ref{ex:zigzag-new-range}, thirteen tail symbols are
read directly. The two remaining symbols follow from
\[
B_2[5]=y_{0,3}+B_1[8],\qquad
B_1[9]=y_{1,3}+B_2[6].
\]
Hence the $15\times18$ tail matrix has row rank $15$, and the
tail is known by time $21$. The short-burst delays
$(25,22,22,10)$ pair with the long-burst delays
$(12,15,15,27)$ to give $37$ for all four groups.
The code attains $9/13$ in either link order, outside the SR range.

The coupled equations illustrate the finite-window recovery in
Lemma~\ref{pw:tail-rank}. Their coefficients are a specialization of
\eqref{pw:coefficient-choice}.
The appendix gives the complete encoder, all eighteen observations,
and both relay maps.

\section{Construction Ranges and Their Contribution}
\label{sec:large-delay}
We compare the exact non-adaptive attainment ranges in
Table~\ref{tab:comparison}. The baseline combines the divisible
regime of Singhvi et al.~\cite{9834645} with the layered construction
of Li et al.~\cite{li2025rateoptimalstreamingcodesthreenode}.
For $u>v$, the latter uses the integer condition
\begin{equation}
 T-u\ge v\left\lceil\frac{T-v}{u}\right\rceil.
 \label{secvi:letter-condition}
\end{equation}
With $T-v=pu+q$, its two ordinary profiles are
$((p+1)v\mathbf1_q,(jv\mathbf1_u)_{j=p,\ldots,1})$
and $(u\mathbf1_q,((ju+q)\mathbf1_u)_{j=1,\ldots,p})$;
the latter is the common RD profile in Lemma~\ref{RD}.
The threshold $T\ge u+v+uv/(u-v)$ is sufficient for
\eqref{secvi:letter-condition}, since it implies
$(T-u)/v\ge(T-v)/u+1$. The integer condition can be weaker:
$(b_1,b_2,T)=(2,3,7)$ satisfies it but not the threshold.

\begin{proposition}[Exact pairing condition for the layered family]
\label{secvi:fixed-family}
The two ordinary component families described above admit a pairing
within delay $T$ under some fixed coordinate permutation if and only
if \eqref{secvi:letter-condition} holds.
\end{proposition}
The proof is given in Appendix~\ref{app:fixed-family}. The sufficient
integer condition is already used in~\cite{li2025rateoptimalstreamingcodesthreenode};
Proposition~\ref{secvi:fixed-family} establishes its necessity for
these fixed component families under every fixed coordinate matching.
It does not characterize all ordinary coordinate-forwarding codes.
For example, the layered profiles fail at $(4,5,20)$: the additional
coordinate requires delay $16$, while every coordinate on the other
hop requires at least $5$. Example~\ref{secvi:p3-example} attains the
bound by changing the component encoder and its delay distribution.

The layered and SR ranges are incomparable: $(1,3,6)$ belongs only
to the former, while $(3,4,12)$ belongs only to the latter.
The SR range is contained in the SR-2 range. In the nondivisible
domain $u>v$, $p\ge2$, and $0<q<u$, write $A=(p-1)(u-v)$.
The layered condition is equivalent to $A+q\ge v$.
Putting $m=u-q$ and $h=v-A-q$, this is $h\le0$, where
prefix recovery suffices. SR-2 also permits $h>0$ when
$qh\le mA$, equivalently $q(v-q)\le uA$.
Thus the final attainment region is the union of the baseline
and SR-2, with SR recorded separately to show the successive additions.
Both extended-profile constructions use $\mathbb F_2$, dimensions
$(k,n)=(T-v,T-v+u)$, and component memory at most $T-v$.

\subsection{Attainment Coverage and Ordinary-Forwarding Limits}
\label{sec:final-coverage}
SR-2 enlarges the SR range for arbitrarily large parameters.
For every integer $d\ge1$, consider
\[
(v,u,T)=(2d+2,3d+2,9d+7).
\]
Here $q=d+1>A=d$, while
$q(v-q)=(d+1)^2\le d(3d+2)=uA$.
Theorem~\ref{pw:achievability} therefore attains the rate upper bound
for both link orders, although the SR condition fails.

The ordinary-forwarding limitations retained in this paper concern
the following two regions:
\begin{align}
 \mathcal C_{\rm low}
 &=\{(b_1,b_2):0<T-u-v<v\},\label{eq:coverage-low}\\
 \mathcal C_{\rm eq}
 &=\{(b,b):b\nmid(T-2b)\}.\label{eq:coverage-equal}
\end{align}
Both sets are restricted to $b_1,b_2\ge1$ and $b_1+b_2\le T$.
Theorem~\ref{secvi:theorem20} excludes exact attainment by the
systematic ordinary class in Definition~\ref{def:ordinary-class}
on $\mathcal C_{\rm low}$. Theorem~\ref{thm:equal-ordinary-allp}
gives a quantitative positive rate gap for that class on
$\mathcal C_{\rm eq}$. Both results permit nonlinear components
and predetermined time variation. These are limitations of ordinary
coordinate forwarding; they do not exclude general non-adaptive relaying.

Table~\ref{tab:coverage} counts ordered burst pairs. Its SR additions
exclude the baseline, and its SR-2 additions exclude both preceding
sets. At $T=80$, the baseline covers $2011$ pairs; SR adds $68$
and SR-2 adds a further $22$, giving $2101$ attained pairs.
The union $\mathcal C_{\rm low}\cup\mathcal C_{\rm eq}$ contains
$1019$ pairs and is disjoint from this attained set. The remaining
$40$ pairs are covered by neither the displayed constructions nor
these two converse regions. These counts make no impossibility
claim for the remaining set.

\begin{table*}[!htbp]
\centering\small
\caption{Construction coverage and ordinary-forwarding exclusions for ordered burst pairs}
\label{tab:coverage}
\begin{tabular}{|r|r|r|r|r|r|r|r|}\hline
$T$&Total&Baseline&\shortstack{SR\\additions}&\shortstack{SR-2\\additions}&\shortstack{Final\\attainment}&\shortstack{Ordinary\\excluded}&Remaining\\\hline
20 & 190 & 131 & 4 & 2 & 137 & 53 & 0\\
40 & 780 & 511 & 14 & 8 & 533 & 241 & 6\\
60 & 1770 & 1139 & 40 & 10 & 1189 & 561 & 20\\
80 & 3160 & 2011 & 68 & 22 & 2101 & 1019 & 40\\
100 & 4950 & 3134 & 94 & 44 & 3272 & 1610 & 68\\
\hline\end{tabular}
\par\smallskip
\begin{minipage}{0.97\textwidth}\footnotesize
The domain is $b_1,b_2\ge1$ and $b_1+b_2\le T$.
The baseline combines the divisible and layered regimes.
Writing the baseline, SR, and SR-2 sets as $\mathcal B,\mathcal S,\mathcal S_2$,
the additions are $\mathcal S\setminus\mathcal B$ and
$\mathcal S_2\setminus(\mathcal B\cup\mathcal S)$.
The ordinary-excluded column counts the union of
\eqref{eq:coverage-low} and \eqref{eq:coverage-equal}, without double counting;
its scope is Definition~\ref{def:ordinary-class}.
The remaining column counts pairs in neither the final attainment set
nor this union. The checks in the source package reproduce all rows
and the $T=80$ point sets.
\end{minipage}
\end{table*}

\begin{inlinefigure}
\centering
\includegraphics[width=0.94\textwidth]{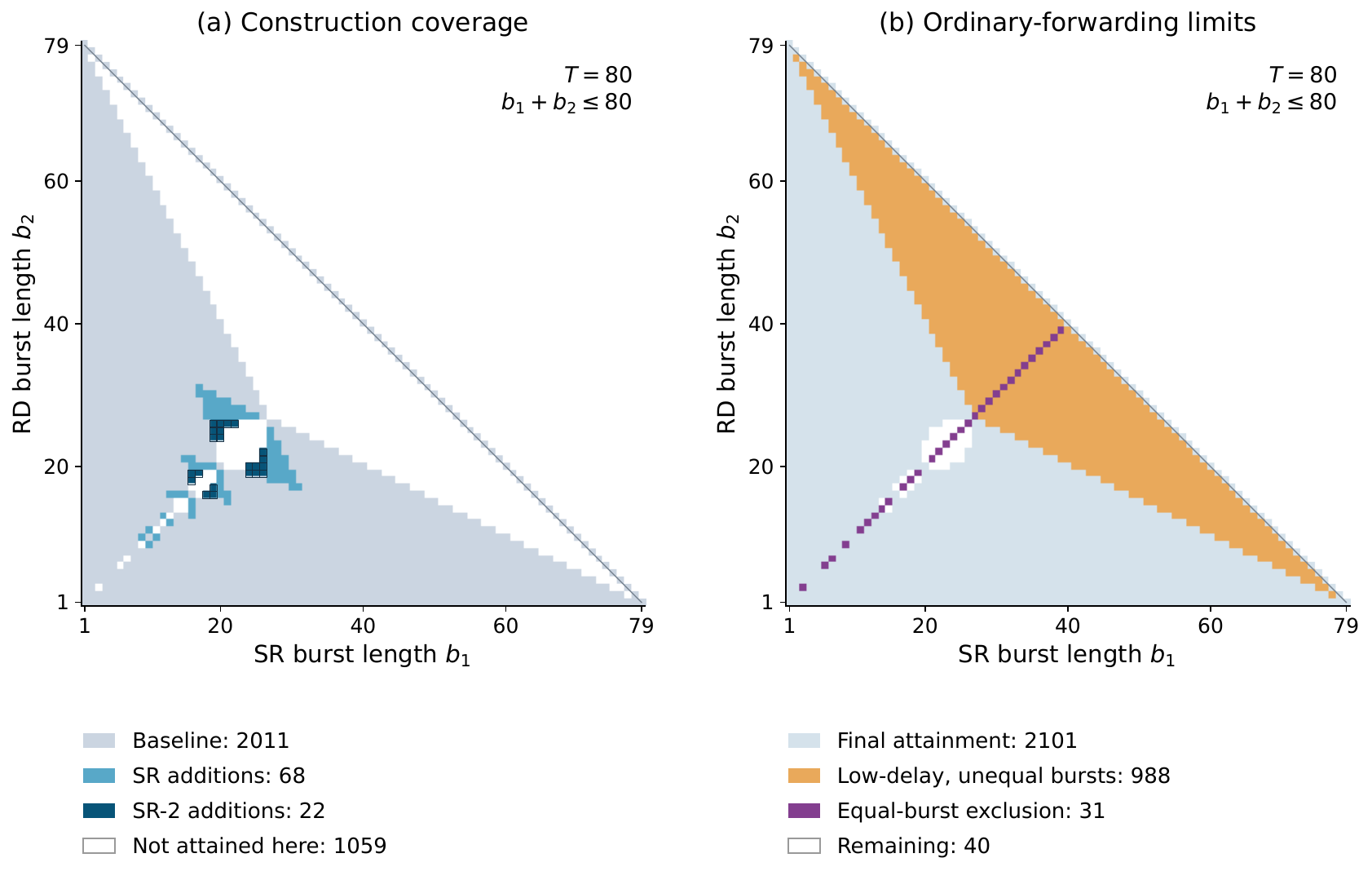}
\caption{Ordered burst-pair coverage at $T=80$.
The left panel shows the baseline and the successive SR and SR-2
additions. The right panel shows the two ordinary-forwarding converse
regions in \eqref{eq:coverage-low}--\eqref{eq:coverage-equal}:
$988$ unequal-burst pairs in the low-delay region and $31$ nondivisible
equal-burst pairs. Equal-burst pairs are assigned to the equal-burst
category when the two regions overlap. The $2101$ attained pairs
and the remaining $40$ pairs are also shown to make the partition
explicit. The exclusions concern Definition~\ref{def:ordinary-class},
not the full class of non-adaptive relay codes.}
\label{fig:parameter_comparison}
\end{inlinefigure}

\begin{samepage}
\section{Worked Examples}\label{sec:examples}
\subsection{Example 1: Extended SR and Ordinary RD}\label{ex:new-only}
Consider $(b_1,b_2,T)=(3,4,12)$ with rate upper bound $9/13$.
The extended-profile condition holds with equality, while
the large-delay integer condition fails:
$T-u=8<3\lceil9/4\rceil=9$.
This example illustrates how the extended profile permits attainment
outside the layered baseline range.

\end{samepage}

Set $(k,n)=(9,13)$ and initialize all negative-time source symbols
to zero. Write $P[t]=(p_1[t],\ldots,p_4[t])$.
\subsubsection{The SR code}
The source encoder has parity equations
\begin{align*}
p_1[t]&=s_6[t-3]+s_5[t-6]+s_1[t-7],&
p_2[t]&=s_7[t-3]+s_5[t-4]+s_2[t-7],\\
p_3[t]&=s_8[t-3]+s_5[t-5]+s_3[t-7],&
p_4[t]&=s_9[t-3]+s_4[t-7].
\end{align*}
Equivalently, $P_3$ has $I_4$ in its last four rows, $P_7$
has $I_4$ in its first four rows, and
$P_4(5,2)=P_5(5,3)=P_6(5,1)=1$.
All remaining entries and blocks are zero.

For a burst at $0,1,2$, subtract the known systematic contributions.
At delay $3$, the last four parity positions reveal
$s_6[i]$, $s_7[i]+s_5[i-1]$, $s_8[i]+s_5[i-2]$, and $s_9[i]$.
At time $6$, the first three parity coordinates recover
$s_5[0],s_5[2],s_5[1]$, respectively.
The first four source coordinates are then directly recovered
at times $i+7$.
The uniform guarantees follow from the SR construction in
Lemma~\ref{delay SR} and its received-or-erased check.
The resulting innovations are
\[
(s_1[t],\ldots,s_6[t],
s_7[t]+s_5[t-1],s_8[t]+s_5[t-2],s_9[t])
\]
with extended profile $(7,7,7,7,6,3,3,3,3)$.

\subsubsection{The RD code and relay map}
The relay sends the prescribed input
\begin{align*}
R[t]=\big(&s_1[t-7],\ldots,s_4[t-7],s_5[t-6],s_6[t-3],\\
&s_7[t-3]+s_5[t-4],s_8[t-3]+s_5[t-5],s_9[t-3]\big).
\end{align*}
These values are available at the scheduled time even when the
relay does not yet know the individual terms in an early mixture.
Its four-burst RD component has parity
\begin{align*}
p'_1[t]&=r_1[t-4]+r_2[t-5]+r_6[t-9],&
p'_2[t]&=r_1[t-1]+r_3[t-5]+r_7[t-9],\\
p'_3[t]&=r_1[t-2]+r_4[t-5]+r_8[t-9],&
p'_4[t]&=r_1[t-3]+r_5[t-5]+r_9[t-9].
\end{align*}
Lemma~\ref{RD} gives ordinary RD profile
$(4,5,5,5,5,9,9,9,9)$.
For instance, the seventh innovation generated at source time $i$
is delivered by $i+12$. Its term $s_5[i-1]$ has already been
delivered by $i-1+6+5=i+10$, so the destination subtracts it.
The eighth coordinate is handled similarly.
Every source coordinate therefore meets deadline $i+12$.
\seventhfix{Figure~\ref{fig:table_r} gives the corresponding relay transmissions in time order.}

\begin{inlinefigure}
\centering
\begingroup
\small
\setlength{\tabcolsep}{3pt}
\renewcommand{\arraystretch}{1.08}
\fitgraphic{0.97\textwidth}{%
\begin{tabular}{|c|c|c|c|c|c|c|c|c|c|c|}
\hline
Symbol / time & 3 & 4 & 5 & 6 & 7 & 8 & 9 & 10 & 11 & 12\\\hline
$r_1[t]$ & $0$ & $0$ & $0$ & $0$ & $\textcolor{red}{s_1[0]}$ & $s_1[1]$ & $s_1[2]$ & $s_1[3]$ & $s_1[4]$ & $s_1[5]$ \\\hline
$r_2[t]$ & $0$ & $0$ & $0$ & $0$ & $\textcolor{red}{s_2[0]}$ & $s_2[1]$ & $s_2[2]$ & $s_2[3]$ & $s_2[4]$ & $s_2[5]$ \\\hline
$r_3[t]$ & $0$ & $0$ & $0$ & $0$ & $\textcolor{red}{s_3[0]}$ & $s_3[1]$ & $s_3[2]$ & $s_3[3]$ & $s_3[4]$ & $s_3[5]$ \\\hline
$r_4[t]$ & $0$ & $0$ & $0$ & $0$ & $\textcolor{red}{s_4[0]}$ & $s_4[1]$ & $s_4[2]$ & $s_4[3]$ & $s_4[4]$ & $s_4[5]$ \\\hline
$r_5[t]$ & $0$ & $0$ & $0$ & $\textcolor{red}{s_5[0]}$ & $s_5[1]$ & $s_5[2]$ & $s_5[3]$ & $s_5[4]$ & $s_5[5]$ & $s_5[6]$ \\\hline
$r_6[t]$ & $\textcolor{red}{s_6[0]}$ & $s_6[1]$ & $s_6[2]$ & $s_6[3]$ & $s_6[4]$ & $s_6[5]$ & $s_6[6]$ & $s_6[7]$ & $s_6[8]$ & $s_6[9]$ \\\hline
$r_7[t]$ & $\textcolor{red}{s_7[0]}$ & $\begin{array}{c}s_7[1]\\+\textcolor{red}{s_5[0]}\end{array}$ & $\begin{array}{c}s_7[2]\\+s_5[1]\end{array}$ & $\begin{array}{c}s_7[3]\\+s_5[2]\end{array}$ & $\begin{array}{c}s_7[4]\\+s_5[3]\end{array}$ & $\begin{array}{c}s_7[5]\\+s_5[4]\end{array}$ & $\begin{array}{c}s_7[6]\\+s_5[5]\end{array}$ & $\begin{array}{c}s_7[7]\\+s_5[6]\end{array}$ & $\begin{array}{c}s_7[8]\\+s_5[7]\end{array}$ & $\begin{array}{c}s_7[9]\\+s_5[8]\end{array}$ \\\hline
$r_8[t]$ & $\textcolor{red}{s_8[0]}$ & $s_8[1]$ & $\begin{array}{c}s_8[2]\\+\textcolor{red}{s_5[0]}\end{array}$ & $\begin{array}{c}s_8[3]\\+s_5[1]\end{array}$ & $\begin{array}{c}s_8[4]\\+s_5[2]\end{array}$ & $\begin{array}{c}s_8[5]\\+s_5[3]\end{array}$ & $\begin{array}{c}s_8[6]\\+s_5[4]\end{array}$ & $\begin{array}{c}s_8[7]\\+s_5[5]\end{array}$ & $\begin{array}{c}s_8[8]\\+s_5[6]\end{array}$ & $\begin{array}{c}s_8[9]\\+s_5[7]\end{array}$ \\\hline
$r_9[t]$ & $\textcolor{red}{s_9[0]}$ & $s_9[1]$ & $s_9[2]$ & $s_9[3]$ & $s_9[4]$ & $s_9[5]$ & $s_9[6]$ & $s_9[7]$ & $s_9[8]$ & $s_9[9]$ \\\hline
$p^{\prime}_1[t]$ & $0$ & $0$ & $0$ & $0$ & $0$ & $0$ & $0$ & $0$ & $\textcolor{red}{s_1[0]}$ & $\begin{array}{c}s_1[1]\\+\textcolor{red}{s_6[0]}\\+\textcolor{red}{s_2[0]}\end{array}$ \\\hline
$p^{\prime}_2[t]$ & $0$ & $0$ & $0$ & $0$ & $0$ & $\textcolor{red}{s_1[0]}$ & $s_1[1]$ & $s_1[2]$ & $s_1[3]$ & $\begin{array}{c}s_1[4]\\+\textcolor{red}{s_7[0]}\\+\textcolor{red}{s_3[0]}\end{array}$ \\\hline
$p^{\prime}_3[t]$ & $0$ & $0$ & $0$ & $0$ & $0$ & $0$ & $\textcolor{red}{s_1[0]}$ & $s_1[1]$ & $s_1[2]$ & $\begin{array}{c}s_1[3]\\+\textcolor{red}{s_8[0]}\\+\textcolor{red}{s_4[0]}\end{array}$ \\\hline
$p^{\prime}_4[t]$ & $0$ & $0$ & $0$ & $0$ & $0$ & $0$ & $0$ & $\textcolor{red}{s_1[0]}$ & $\begin{array}{c}s_1[1]\\+\textcolor{red}{s_5[0]}\end{array}$ & $\begin{array}{c}s_1[2]\\+s_5[1]\\+\textcolor{red}{s_9[0]}\end{array}$ \\\hline
\end{tabular}}
\endgroup
\caption{\seventhfix{Relay transmission for $(b_1,b_2,T)=(3,4,12)$ in Example 1.
Entries expand the relay inputs and parity symbols into source symbols.
Red terms belong to $S[0]$. The extended SR and ordinary RD profiles
give coordinate delay sums at most $12$.}}
\label{fig:table_r}
\end{inlinefigure}

\subsection{Example 2: Ordinary SR and Extended RD}
Now interchange the link parameters:
$(b_1,b_2,T)=(4,3,12)$, again with rate $9/13$.
The SR encoder is the four-burst component of Example 1,
with $r_i$ replaced by $s_i$ in its parity equations.
It has ordinary profile $(4,5,5,5,5,9,9,9,9)$.
The RD encoder is the three-burst component of Example 1,
with $s_i$ replaced by its own inputs $r_i$.
Its extended profile is $(7,7,7,7,6,3,3,3,3)$.

The relay uses the separability in Definition~\ref{def:eds}:
\begin{align*}
R[t]=\big(&s_1[t-4],s_2[t-5],s_3[t-5],s_4[t-5],s_5[t-5],\\
&s_6[t-9],s_7[t-9]+s_5[t-6],
s_8[t-9]+s_5[t-7],s_9[t-9]\big).
\end{align*}
All required source coordinates and past relay inputs are available.
The RD decoder recovers
\[
r_7[t]+r_5[t-1]=s_7[t-9],\qquad
r_8[t]+r_5[t-2]=s_8[t-9]
\]
within delay $3$. Hence these source coordinates meet delay
$9+3=12$. The other coordinate delay sums are
$4+7$, $5+7$, $5+6$, or $9+3$, all at most $12$.
The past-term cancellation is performed before RD encoding.
\seventhfix{The transmission table follows analogously to Fig.~\ref{fig:table_r}
by substituting this relay map into the RD parity equations.}

\subsection{Example 3: Ordinary Profiles on Both Hops}
For $(b_1,b_2)=(2,3)$, the exact integer condition of the
large-delay family gives $T=5$ and every $T\geq7$.
Theorem~\ref{secvi:theorem20} excludes ordinary coordinate
forwarding at $T=6$.
We now write the $(2,3,7)$ rate-$5/8$ code explicitly.

\subsubsection{The SR component}
The SR parity equations are
\[
p_1[t]=s_3[t-2]+s_1[t-4],\quad
p_2[t]=s_4[t-2]+s_2[t-4],\quad p_3[t]=s_5[t-2].
\]
The nonzero blocks are
\[
P_2=\begin{pmatrix}0_{2\times3}\\I_3\end{pmatrix},
\qquad
P_4=\begin{pmatrix}I_2&0_{2\times1}\\
\multicolumn{2}{c}{0_{3\times3}}\end{pmatrix}.
\]
For a burst at $0,1$, the effective matrix is
\[
\mathbf P_{\rm SR}=
\begin{pmatrix}P_2&P_3&P_4&0\\P_1&P_2&P_3&P_4\end{pmatrix},
\qquad P_1=P_3=0.
\]
The last three coordinates are recovered at delay $2$;
the first two at delay $4$. Thus the ordinary SR profile is
$(4,4,2,2,2)$. 

\subsubsection{The RD component and pairing}
The three-burst RD parity equations are
\begin{align*}
p'_1[t]&=r_1[t-3]+r_3[t-5],\\
p'_2[t]&=r_2[t-3]+r_4[t-5],\\
p'_3[t]&=r_1[t-1]+r_2[t-2]+r_5[t-5].
\end{align*}
The ordinary RD profile is $(3,3,5,5,5)$.
For a burst at $0,1,2$, the parity packets at $3,4$ recover
all six erased symbols from the first two coordinates:
$p'_1[3],p'_2[3]$ give $r_1[0],r_2[0]$;
$p'_1[4],p'_2[4]$ give $r_1[1],r_2[1]$;
$p'_3[4]$ gives $r_2[2]$ after subtracting a known
systematic input, and then $p'_3[3]$ gives $r_1[2]$.
The final three coordinates are directly recovered at delay $5$.

The relay forwards
\[
R[t]=(s_1[t-4],s_2[t-4],s_3[t-2],s_4[t-2],s_5[t-2]).
\]
The paired sums are $4+3=2+5=7$, so this ordinary
coordinate-forwarding code meets the deadline and attains $5/8$.
\seventhfix{Its transmission table is obtained by the same substitution
as in Fig.~\ref{fig:table_r}.}

\setcounter{example}{3}
\section{\reviewfix{Limits of Systematic Ordinary Coordinate Forwarding}}
\label{sec:ordinary-forwarding}

Ordinary coordinate forwarding must recover a source coordinate
before sending its delayed value to the next hop. At the rate bound,
some coordinates on the longer-burst hop necessarily use the full
delay budget. Their partners must then be recovered early on the
other hop. We use this requirement to prove the low-delay converse,
and count the entire delay distribution for equal bursts.
\eighthfix{The supporting point-to-point lemmas are proved in
Appendix~\ref{app:ordinary-proofs}.}

\subsection{Ordinary Coordinate Forwarding}

\begin{definition}[Systematic ordinary coordinate forwarding]\label{def:ordinary-class}
Both hops use systematic causal components of common dimensions
$(k,n=k+r)$, where $r$ is the number of parity symbols.
Their encoding rules are fixed before transmission, independently
of erasures, and may vary with time.
An \emph{ordinary coordinate-forwarding code} uses a relay input map
\begin{equation}
 R_{\pi(i)}[t]=s_i[t-a_i],\qquad
 a_i+d_{\pi(i)}\leq T,\quad i\in[k].
 \label{secvi:ordinary-map}
\end{equation}
Here $\pi$ is any fixed permutation of $[k]$, chosen jointly
with the two components and their delay allocation.
The delay $a_i$ concerns SR source coordinate $i$;
$d_j$ concerns RD input coordinate $j$.
Thus the relay may send the slowest SR coordinate to the fastest
RD position. There is no requirement that coordinate $i$ be
paired with RD position $i$. Each component must
provide its ordinary coordinate deadlines for every input stream.
In particular, $a_i\geq b_1$ and $d_i\geq b_2$, so the full component
deadlines are at most $T-b_2$ and $T-b_1$, respectively.
All deadlines and relay shifts are uniform over source time.
The component encoders may be nonlinear and predetermined time
varying unless a result explicitly assumes linearity.
\end{definition}

\noindent\emph{Coordinate matching.}
\fourthfix{Let $\widetilde R_i[t]=R_{\pi(i)}[t]$.
The RD delay assigned to this relabelled input is
$\widetilde d_i=d_{\pi(i)}$, where $d_j$ denotes the original
delay of RD position $j$.}
Permuting the systematic positions within every RD packet preserves
packet erasures, dimensions, and all decoding guarantees.
It also preserves linearity and time invariance when imposed.
Thus the relay equation becomes
$\widetilde R_i[t]=s_i[t-a_i]$ without restricting the matching.

For two given delay multisets, let
$a_{(1)}\leq\cdots\leq a_{(k)}$ and
$d_{(1)}\leq\cdots\leq d_{(k)}$ be their sorted entries.
A feasible permutation exists exactly when
\begin{equation}
 a_{(i)}+d_{(k+1-i)}\leq T,\qquad i\in[k].
 \label{eq:opposite-delay-matching}
\end{equation}
To see necessity, if the $i$th inequality fails, the $k-i+1$
SR entries at least $a_{(i)}$ have only $k-i$ RD entries
small enough to pair with them. Sufficiency follows by pairing
in the displayed opposite order.
We allow this optimal matching throughout.

The layered construction in Section~\ref{sec:large-delay} belongs
to this class. Our extended construction uses the same component
restrictions but forwards innovations in place of some original
coordinates. In the converses, systematic packets let us isolate
the parity information, and common dimensions give the same
$r=n-k$ on both hops. 

Write $u=\max\{b_1,b_2\}$ and $v=\min\{b_1,b_2\}$. To use one
parameterization for both link orders, set
\[
 T-v=pu+q,\qquad 0\leq q<u.
\]
When $b_1=v<b_2=u$, this is precisely $T-b_1=pb_2+q$.
The rate upper bound is
\[
 R^\star=\frac{T-v}{T-v+u}.
\]
At equality, $k/r=(T-v)/u=p+q/u$.

\subsection{The Low-Delay Converse}

\ifdefined\ReadableConverseStandalone\newpage\fi
\begin{theorem}[Low-delay converse]
\label{secvi:theorem20}
Consider the systematic ordinary coordinate-forwarding class of
Definition~\ref{def:ordinary-class}, allowing nonlinear and predetermined
time-varying component encoders.
If $0<T-u-v<v$, no such code attains $R^\star$.
\end{theorem}
\begin{eighthblock}
\begin{proof}
\noindent\emph{1) Capacity forces late coordinates.}
Write $T=u+v+q$, $0<q<v$. Attaining the bound requires
\begin{equation}
 \frac{k}{r}=\frac{u+q}{u}.
 \label{secvi:p1-rate}
\end{equation}
The $u$-burst component has full delay at most $D_\ell=T-v=u+q$.
Its actual maximum minimum guaranteed delay equals $D_\ell$. Indeed,
a smaller $D'$ would give
$k/n=D_\ell/(D_\ell+u)>D'/(D'+u)$, above point-to-point capacity.
Lemma~\ref{secvi:late-coordinates} therefore forces at least
$r$ ordinary coordinates to need delay $D_\ell$.

\noindent\emph{2) Any reordering forces an early group.}
A permutation pairs those coordinates with at least $r$
distinct coordinates on the $v$-burst hop, each due by
$T-D_\ell=v$. If the $u$-burst hop is SR, use the image under
$\pi$; if it is RD, use the inverse image.
The count is unchanged by the chosen matching.

\noindent\emph{3) The other hop has too little parity.}
Its full delay is at most $T-u=v+q$.
Lemma~\ref{secvi:early-count}, with $m\geq r$, gives
$k+qr\leq(q+1)r$, or $k\leq r$.
This contradicts \eqref{secvi:p1-rate}, since $q>0$.
\end{proof}
\end{eighthblock}

\subsection{Equal Bursts: A Converse for Every $p$}
\label{sec:permutation-counts}
\begin{eighthblock}
For equal bursts, we count the ordinary delay distribution on both
hops. Let $F_\ell(h)$ count the coordinates on hop $\ell$ with
guaranteed delay at most $h$. Lemma~\ref{lem:ordinary-prefix-count}
gives
\[
 \sum_{\tau=0}^{b-1}F_\ell(h-\tau)\leq(h-b+1)r,
 \qquad h\ge b,\quad \ell\in\{1,2\}.
\]
At suitable windows, every matched coordinate pair contributes equally to the sum.
\end{eighthblock}

This count uses the independence of the original input coordinates.
An innovation that contains unrecovered past symbols need not
satisfy the same count.

\begin{theorem}[An all-$p$ equal-burst converse]
\label{thm:equal-ordinary-allp}
Let $b_1=b_2=b$, $T\geq2b$, and write
$T-b=pb+q$, $p\geq1$, $0\leq q<b$.
Every code in the systematic ordinary coordinate-forwarding class
of Definition~\ref{def:ordinary-class}, allowing any fixed permutation and predetermined,
possibly nonlinear and time-varying components of common dimensions,
satisfies
\begin{equation}
 R\leq R_{\rm ord}:=
 \frac{bp(p+1)}{b(p+1)^2-q}.
 \label{eq:equal-ordinary-rate-bound}
\end{equation}
For $q>0$, this is strictly smaller than
$R^\star=(pb+q)/((p+1)b+q)$, with
\begin{equation}
 R^\star-R_{\rm ord}
 =\frac{q(b-q)}
 {((p+1)b+q)(b(p+1)^2-q)}>0.
 \label{eq:equal-ordinary-gap}
\end{equation}
\reviewfix{Consequently systematic ordinary coordinate forwarding
of common hop dimensions in Definition~\ref{def:ordinary-class}
attains $R^\star$ if and only if}
$b\mid(T-2b)$. The claim includes every $p\geq3$.
\end{theorem}
\begin{eighthblock}
\begin{proof}
\noindent\emph{1) Count matched delay pairs.}
Pair the two profiles by any feasible permutation.
If one member has guaranteed delay $a$, the other has delay
at most $T-a$; enlarge the latter guarantee to $T-a$.
This only relaxes its deadline. Let $N_a$ count these pairs,
so $b\leq a\leq T-b=pb+q$ and $\sum_aN_a=k$.
Write $[z]_0^b=\min\{b,\max\{0,z\}\}$.
A pair contributes $[h-a+1]_0^b+[h-T+a+1]_0^b$
to the two burst-prefix counts at $h$.

\noindent\emph{2) Choose windows with a constant contribution per pair.}
Apply Lemma~\ref{lem:ordinary-prefix-count} on both hops at
$h_j=(j+1)b-1$, $j=1,\ldots,p$, and add the inequalities.
The elementary identity
$\sum_{j=0}^{p-1}[x-jb]_0^b=x$ for $0\leq x\leq pb$
partitions an interval of length $x$ into blocks of width $b$.
Apply it with $x=(p+1)b-a$ and $x=a-q$.
For every allowed $a$, this gives
\[
\begin{split}
 \sum_{j=1}^p[(j+1)b-a]_0^b&=(p+1)b-a,\\
 \sum_{j=1}^p[a-(p-j)b-q]_0^b&=a-q.
\end{split}
\]
Hence \emph{every matched pair}, regardless of its position or
delay, contributes exactly $(p+1)b-q$ to the sum.
The two hops together provide at most
$2r\sum_{j=1}^p(h_j-b+1)=bp(p+1)r$.
Therefore
\[
 ((p+1)b-q)k\leq bp(p+1)r.
\]
\noindent\emph{3) Convert the count into a rate gap.}
Converting $k/r$ to $R=k/(k+r)$ proves
\eqref{eq:equal-ordinary-rate-bound}; direct subtraction gives
\eqref{eq:equal-ordinary-gap}.
For $q=0$, the bound equals $p/(p+1)=R^\star$; the divisible
case of Theorem~\ref{thm:extended-achievability} attains it
by ordinary identity-tap forwarding.
Finally, $q=0$ is equivalent to $b\mid(T-2b)$.
\end{proof}
\end{eighthblock}
\begin{revisionblock}
\begin{reviewblock}
The bound $R_{\rm ord}$ applies uniformly over the finite common
packet dimensions of Definition~\ref{def:ordinary-class}, including
nonlinear and predetermined time-varying components. It therefore
bounds the rate supremum of that entire systematic ordinary class
strictly below $R^\star$ when $q>0$. Tightness of $R_{\rm ord}$
in this case remains open. General fixed-rate non-adaptive attainment
of $R^\star$ for equal bursts and $q>0$ also remains open.
\end{reviewblock}
\end{revisionblock}

For example, $(b_1,b_2,T)=(4,4,17)$ has $p=3,q=1$ and
$R^\star=13/17$, whereas ordinary forwarding has
$R\leq16/21<13/17$. At $(4,4,21)$, $p=4,q=1$ and
$R\leq80/99<17/21=R^\star$.
Both bounds allow arbitrary fixed matching. The equal-burst
assumption matters: the ordinary code at $(4,5,20)$ uses
unequal bursts and lies outside this theorem's scope.

For $(6,6,20)$, the ordinary bound is $9/13<7/10=R^\star$.
Whether a more general fixed-rate non-adaptive relay code attains
$7/10$ remains open. The adaptive construction of \cite{10764771}
approaches the same target $(T-b)/T$ asymptotically, with relay
information sets depending on the SR burst.
The target rate is the same; the available relay operations and
exact versus asymptotic attainment differ.

\subsection{Why the Layered Condition Is Not Necessary for All Ordinary Codes}

\begin{example}[Ordinary forwarding outside the large-delay region]
\label{secvi:p3-example}
There is a binary, systematic, time-invariant ordinary
coordinate-forwarding code for $(b_1,b_2,T)=(4,5,20)$
with $(k,n)=(16,21)$.
It attains the upper bound $16/21$, even though
$T-b_1=16=3\cdot5+1$ and
$T-u=15<4\lceil16/5\rceil=16$.
\end{example}
The old layered allocation leaves too little time for its extra
coordinate. The redesigned component instead moves that coordinate
one slot earlier and uses spare delay on other groups. Appendix
\ref{app:ordinary-counterexample} gives its binary encoder, decoder,
and the scaled family $(4\lambda,5\lambda,20\lambda)$.
This example changes the component profile itself; a permutation
of the old profile cannot achieve the same effect.

\section{Conclusion}\label{sec:conclusion}
SR-2 enlarges the binary attainment condition from
$q\le(p-1)(u-v)$ to $q(v-q)\le(p-1)u(u-v)$ using the common
ordinary long-burst component and fixed finite packet dimensions.
Finite-window tail recovery meets each coordinate deadline and
supports both link orders.

The two converses concern systematic ordinary coordinate forwarding
of common hop dimensions, with fixed coordinate matching and
source-time-uniform relay shifts and deadlines. Such codes cannot
attain the rate upper bound when $0<T-u-v<v$. For equal bursts,
the all-$p$ bound gives a positive rate gap in every nondivisible
case, and ordinary forwarding attains the bound exactly when
$b\mid(T-2b)$. Both converses allow nonlinear components and
predetermined time variation. Tightness of the equal-burst upper
bound, a complete characterization of ordinary-forwarding attainment,
and the full non-adaptive attainment region remain open.

\appendices
\section{SR-2 Block Entries and Detailed Decoding}
\label{app:sr2}
This appendix verifies the recovery blocks and deadlines of the
SR-2 encoder assigned in Section~\ref{sec:SR2construction}.
Use the parameters and source-group order of that section.
The main parity recovers a prefix of length $\min\{L,v\}$.
If $h>0$, the remaining $h$ packets form the tail, and
\begin{equation}
q(v-q)\le uA\quad\Longleftrightarrow\quad qh\le mA.
\label{pw:count}
\end{equation}

\subsection{Parity Expansions and Recovery-Block Entries}
For the decoding calculation, split the parity produced by
\eqref{pw:short-taps} into its first $q$ and last $m$ coordinates,
denoted by $P_F[t]$ and $P_A[t]$. Expanding the assigned rows gives
\begin{equation}
 \begin{split}
 P_F[t]&=F[t-D]+B[t-pv]+U_F[t-\delta]
                         +\sum_{j=1}^{p-2}G_{j,F}[t-jv],\\
 P_A[t]&=A^{\rm o}[t-E]+U_A[t-\delta]
       +\sum_{j=1}^{p-2}G_{j,A}[t-jv]
       +\sum_{j=0}^{d}B[t-\lambda_j]C_j^{\mathsf T}.
 \end{split}
 \label{pw:source-parities}
\end{equation}
The subscripts $F,A$ split $U$ and each $G_j$ into their first
$q$ and last $m$ coordinates.
Write $\mathbf y_{t,w}=(\bar P[t],\ldots,\bar P[t+w-1])$.
The row and column grouping in~\eqref{pw:windows} gives
\begin{equation}
 \fitmatrix{
 \bigl((\mathbf y_{jv,v})_{j=1}^{p-2},\mathbf y_{\delta,v},
          \mathbf y_{pv,A},\mathbf y_{E,q},\mathbf y_{D,v}\bigr)
 =\bigl((\mathbf G_j)_{j=1}^{p-2},\mathbf U,\mathbf B,
                              \mathbf A^{\rm o},\mathbf F\bigr)
                              \widehat P_{\srnamefix{\rm SR\text{-}2}}.}
 \label{pw:sr2-regrouping}
\end{equation}
All remaining blocks are obtained by placing the same assigned
coefficients at their time positions. More explicitly, for
$X\in\{U,C,J,F\}$ use
\[
 (t_X,w_X)=(\delta,v),\ (pv,A),\ (E,q),\ (D,v),
 \quad\text{respectively},
\]
and fill their $q\times u$ small blocks by
\begin{equation}
 (\mathcal B_X)_{i,s}
  =\ind{t_X+s-i=pv}\nz{J_F}
   +\sum_{j=0}^{d}\ind{t_X+s-i=\lambda_j}
                                  \nz{C_j^{\mathsf T}}J_A,
 \quad 0\le i<v,\quad0\le s<w_X.
 \label{pw:sr2-block-entries}
\end{equation}
The $m\times u$ blocks of $\mathcal A_J,\mathcal A_F$ and
the $q\times u$ blocks of $\mathcal F_F$ are
\[
 (\mathcal A_J)_{i,s}=\ind{s=i}\nz{J_A},\qquad
 (\mathcal A_F)_{i,s}=\ind{q+s=i}\nz{J_A},\qquad
 (\mathcal F_F)_{i,s}=\ind{s=i}\nz{J_F},
\]
within the stated row and column ranges. These formulas specify
the remaining blocks of $P_{\srnamefix{\rm SR\text{-}2}}$.

\subsection{\srnamefix{SR-2 Tail Recovery}}

Before time $D$, the first $q$ parity coordinates contain no
erased $F$ or $A^{\rm o}$ symbols. In particular,
\begin{equation}
 \bar P_F[pv+i]=B[i],\qquad 0\le i<\min\{L,v\}.
 \label{pw:prefix}
\end{equation}
If $L\ge v$, this already recovers the whole $B$ group.
Otherwise $h=v-L>0$ and the remaining packets are
$B[L],\ldots,B[v-1]$.

Use the last $m$ parity coordinates in the $A$ packets
$pv,\ldots,E-1$ for this tail. The $A^{\rm o}$ assignment
does not start until $E$, so this window contains only $B$
unknowns. Denote the prefix, tail, and auxiliary observations by
\[
 \begin{gathered}
 \mathbf b_{\rm pre}=(B[0],\ldots,B[L-1]),\quad
 \mathbf b_{\rm tail}=(B[L],\ldots,B[v-1]),\\
 \mathbf y_{\rm aux}=(\bar P_A[pv],\ldots,\bar P_A[E-1]).
 \end{gathered}
\]
Reading the $B$ coefficients of \eqref{pw:short-taps} gives
\begin{equation}
 \begin{gathered}
 \mathbf y_{\rm aux}=\mathbf b_{\rm pre}\mathcal R_{\rm pre}
                            +\mathbf b_{\rm tail}\mathcal M,\\
 \beta=v-\theta,\qquad C_j=0\quad(j\notin[0,d]),\\
 \mathcal R_{\rm pre}=(C_{i-s-\beta}^{\mathsf T})_{
                                  0\le i<L,\,0\le s<A},\qquad
 \mathcal M=(C_{L+i-s-\beta}^{\mathsf T})_{
                                  0\le i<h,\,0\le s<A}.
 \end{gathered}
 \label{pw:tail-matrix}
\end{equation}
The tail is uniquely recoverable when the $qh\times mA$ matrix
$\mathcal M$ has row rank $qh$. Thus the count $qh\le mA$
is necessary for this auxiliary matrix. The binary
choices in~\eqref{pw:coefficient-choice} attain that rank.

Use the coefficient choices and shifts in
\eqref{pw:coefficient-choice}. Every nonzero correction lag
$\theta-j$ is positive.

The tail systems take the simple forms
\begin{equation}
 \begin{array}{c|c|c}
 \text{coefficient choice}&\mathcal R_{\rm pre}&\mathcal M\\ \hline
 m\ge q,\ A\ge h>0&0&(\nz{I_h\otimes C_0^{\mathsf T}}\mid0)\\
 h>A&0&(C_{i-s}^{\mathsf T})_{0\le i<h,\,0\le s<A}\\
 m<q,\ h>0&\text{retain the prefix term}
                  &(C_{d+i-s}^{\mathsf T})_{0\le i<h,\,0\le s<A}
 \end{array}
 \label{pw:tail-forms}
\end{equation}
In the first row, $I_h\otimes C_0^{\mathsf T}$ just means
$h$ copies of $C_0^{\mathsf T}$ on the diagonal.
Each displayed tail matrix has full row rank, as stated in
Lemma~\ref{pw:tail-rank}.
The first two rows recover the tail from the auxiliary window
alone, by $E-1$. In the last row, store the observations and
subtract the prefix term once \eqref{pw:prefix} has given the
whole prefix, by $pv+L-1=D-1$. The tail is then recovered by $D-1$.

\refine{The following subsection proves the rank assertion of
Lemma~\ref{pw:tail-rank} using the actual finite observation window.}

\subsection{\srnamefix{Binary Coefficients and the SR-2 Tail-Rank Proof}}
\label{pw:zigzag}

\reviewfix{The polynomial zigzag matrices and their dual-minimal-basis
degree properties are supplied by~\cite{DeTeranZigzag2016}.
We give the binary assignment explicitly and verify the degree
properties used in the proof. We then apply them to the two finite
observation windows in Lemma~\ref{pw:tail-rank}, including the
prefix-subtraction time needed by the SR-2 delay profile.}

First consider the actual parameters $(v,u,p,q)=(6,7,2,1)$.
Here $A=1$, $L=2$, $h=4$, and $m=6$. The tail is
$B[2],B[3],B[4],B[5]$, and only one auxiliary packet, at
time $pv=12$, is available. Choose
\begin{equation}
 C(z)=\begin{pmatrix}0\\0\\\nz1\\\nz z\\\nz{z^2}\\\nz{z^3}\end{pmatrix},
 \quad\theta=4,\quad(\lambda_0,\lambda_1,\lambda_2,\lambda_3)
                              =(10,9,8,7).
 \label{pw:tail-example}
\end{equation}
The cleaned last six parity positions at time $12$ are exactly
\[
 \bar P_A[12]=(0,0,B[2],B[3],B[4],B[5]).
\]
Thus $\mathcal M=(0_{4\times2}\mid\nz{I_4})$ and the
tail is read directly. The general zigzag rule makes the same
kind of assignment when several auxiliary packets and several
coordinates per $B$ packet must be used.

\noindent\emph{1) Distribute and group the degrees.}
For positive integers $r,c,d$, the matrix $\ZZ(r,c,d)$ has
$r$ rows and $r+c$ columns. Set
\begin{equation}
 a=\left\lfloor\frac{rd}{c}\right\rfloor,\qquad
 t=rd-ac,\qquad
 \varepsilon_j=a+\ind{j\le t},\quad1\le j\le c.
 \label{pw:balanced}
\end{equation}
These integers sum to $rd$ and differ by at most one.
Temporarily remove the zero entries, whose number is $c_0$.
Group the remaining positive entries in order, ending a block
whenever the running sum is divisible by $d$. There are
$N_{\rm blk}$ such blocks. For one block, write its entries
as $\eta_1,\ldots,\eta_s$, put
$\sigma_j=\sum_{\ell=1}^j\eta_\ell$, and put $R=\sigma_s/d$.
This block has size $R\times(R+s)$.

\noindent\emph{2) Fill each block with monomials.}
Use one-based indices within the block and calculate
\[
 b_i=i+\#\{j:\sigma_j<(i-1)d\},\quad1\le i\le R,
 \qquad
 \alpha_j=\left\lfloor\frac{\sigma_j-1}{d}\right\rfloor,
 \quad\omega_j=\sigma_j-d\alpha_j.
\]
Here $1\le\omega_j\le d$, including $\omega_s=d$.
Start with a zero matrix $Z_b$ and make only the assignments
\begin{equation}
 \begin{aligned}
 (Z_b)_{i,b_i}&=\nz1,&&1\le i\le R,\\
 (Z_b)_{i,b_{i+1}}&=\nz{z^d},&&1\le i<R,\\
 (Z_b)_{\alpha_j+1,\,\alpha_j+1+j}
                  &=\nz{z^{\omega_j}},&&1\le j\le s.
 \end{aligned}
 \label{pw:zigzag-fill}
\end{equation}
The proper partial sums $\sigma_j$ are not divisible by $d$,
so these positions do not conflict. Assemble the blocks as
\begin{equation}
 \ZZ(r,c,d)=(0_{r\times c_0}\mid
          \operatorname{diag}(\nz{Z_1},\ldots,\nz{Z_{N_{\rm blk}}})).
 \label{pw:zigzag-assembly}
\end{equation}
The block row counts sum to $r$, and the total column count
is $c_0+r+(c-c_0)=r+c$. Every nonzero coefficient is $1$.
Finally, take the coefficient of $z^j$ in the chosen $C(z)$
to obtain the binary $C_j$ in \eqref{pw:short-taps}.

For example, $\ZZ(1,5,3)$ has balanced degrees
$(1,1,1,0,0)$. Its two zero entries produce two leading zero
columns; the positive block gives $(1,z,z^2,z^3)$. Its transpose
is exactly \eqref{pw:tail-example}. A two-row example is
\[
 \ZZ(2,1,2)=\begin{pmatrix}
 \nz1&\nz{z^2}&0\\0&\nz1&\nz{z^2}
 \end{pmatrix}.
\]
Taking coefficients of successive powers of $z$ gives the binary
matrices used by the encoder.

The following elementary property is enough for both tail cases.
In its proof, vectors are columns, their entries are polynomials,
and their degree is the largest entry degree; $\deg0=-\infty$.

\begin{lemma}[Degree bounds for the zigzag coefficients]
\label{pw:kernel-lemma}
Let $Z=\ZZ(r,c,d)$. It has full row rank. Every nonzero
polynomial solution of $Zx=0$ has degree at least
$\lfloor rd/c\rfloor$. There is an explicitly determined
$c\times(r+c)$ polynomial matrix $K$ such that
\begin{equation}
 KZ^{\mathsf T}=0,\qquad
 \text{each row of }K\text{ has degree at most }
                      \left\lceil\frac{rd}{c}\right\rceil,
 \label{pw:degrees}
\end{equation}
and every nonzero polynomial solution of $Ky=0$ has degree
at least $d$.
\end{lemma}
\begin{proof}
We describe the two sets of solutions directly. In one positive
block, the main columns are $b_1,\ldots,b_R$ and the remaining
columns are $t_j=\alpha_j+1+j$, $1\le j\le s$.
They partition all columns: their order is obtained by merging
$0,d,\ldots,(R-1)d$ with $\sigma_1,\ldots,\sigma_s$.
The main-column submatrix $H$ has diagonal $1$ and
superdiagonal $z^d$, and is therefore polynomially invertible.
This proves full row rank.

\fourthfix{Let $e_i$ be the $i$th unit column of length $R+s$
in this local column ordering.} Define
\[
 f_j=e_{t_j}+\sum_{i=1}^{\alpha_j+1}
                  z^{\sigma_j-(i-1)d}e_{b_i},\qquad
 g_1=f_1,\quad g_j=f_j+z^{\eta_j}f_{j-1}\quad(j\ge2).
\]
Substitution in \eqref{pw:zigzag-fill} shows $Z_bf_j=0$:
the successive main-column terms cancel, and the last one
cancels the term in column $t_j$. Each choice of the free
coordinates gives a unique choice of main coordinates through
$H$. Hence the $f_j$ generate all polynomial solutions of
$Z_bx=0$. The triangular change from $f_j$ to $g_j$ has
diagonal $1$ and a polynomial inverse, so the $g_j$ do too.

The highest-degree term of $g_1$ has degree $\eta_1$ and
is in column $b_1$. In $g_j$ for $j\ge2$, common main-column
terms cancel. The remaining main-column degrees are at most
$\eta_j+\omega_{j-1}-d<\eta_j$, while column $t_{j-1}$
has the term $z^{\eta_j}$. Thus the leading coefficient
vectors of $g_1,\ldots,g_s$ are independent. Consequently
\[
 \deg\Bigl(\sum_j g_j a_j\Bigr)
                 =\max_{a_j\ne0}(\eta_j+\deg a_j).
\]
No nonzero solution can have degree below the smallest
$\eta_j$. Define $K_b=(g_1\ \cdots\ g_s)^{\mathsf T}$;
its row degrees are precisely the $\eta_j$.

It remains to bound solutions of $K_by=0$. Choose
$a=(H^{\mathsf T})^{-1}y_{\rm main}$ and put
$w=y-Z_b^{\mathsf T}a$. Its main coordinates are zero and
$K_bw=0$. The restriction of the $g_j$ to their free
coordinates is triangular with diagonal $1$, so its free
coordinates are zero as well. Hence $y=Z_b^{\mathsf T}a$.
The leading coefficients of the rows of $Z_b$ occur in the
distinct columns $b_2,\ldots,b_R,t_s$, at degree $d$.
They cannot cancel, giving $\deg y=d+\deg a$ for $y\ne0$.

Take direct sums over all blocks. Each prepended zero column
adds a unit solution of degree zero to $K$. Reorder the rows
of $K$ to match the original $\varepsilon_j$ order. Since
$\min_j\varepsilon_j=\lfloor rd/c\rfloor$ and
$\max_j\varepsilon_j=\lceil rd/c\rceil$, all the claimed
bounds follow.
\end{proof}

\begin{proof}[Proof of Lemma~\ref{pw:tail-rank}]
It suffices to show that zero auxiliary observations, after
subtracting any prefix term, force every tail symbol to be zero.
The direct-assignment case already has a diagonal copy of
$C_0^{\mathsf T}=(I_q\mid0)$ for each tail packet, so its
rank is $qh$.

For the two zigzag cases, let $x_i=B[L+i]^{\mathsf T}$ and
list the tail in reverse order as
\[
 X_*(z)=\sum_{i=0}^{h-1}x_i z^{h-1-i},\qquad\deg X_*<h.
\]
The reversal aligns the tail observations with coefficients
of the polynomial product below.

\noindent\emph{Prefix-subtracted case: all coefficients of a
product are observed.}
Here $d=A-h>0$ and $C=\ZZ(m,q-m,d)$. If $y_s$ is the
transpose of auxiliary packet $s$ after prefix subtraction,
the block $C_{d+i-s}^{\mathsf T}$ gives
\[
 \sum_{s=0}^{A-1}y_s z^{A-1-s}=C(z)X_*(z).
\]
The degree of this product is at most $d+h-1=A-1$, so the
$A$ observations contain the entire product. If they are zero,
$CX_*=0$. The preceding lemma gives the degree bound
\[
 \left\lfloor\frac{m(A-h)}{q-m}\right\rfloor\ge h,
\]
where the inequality follows from $mA\ge qh$. Since
$\deg X_*<h$, this forces $X_*=0$.

\noindent\emph{Pure-tail case: the middle coefficients are enough.}
Here $d=h-A>0$ and $C=Z^{\mathsf T}$ with
$Z=\ZZ(q,m-q,d)$. The observations in
$\mathcal M_{i,s}=C_{i-s}^{\mathsf T}$ are exactly the
coefficients of $CX_*$ at degrees $d,\ldots,h-1$.
If these coefficients vanish, the only possible parts left are
\begin{equation}
 CX_*=L_0+z^hH_0,\qquad\deg L_0<d,\quad\deg H_0<d.
 \label{pw:missing-ends}
\end{equation}
The matrix $K$ from the preceding lemma satisfies $KC=0$,
and its row degrees are at most
\[
 \left\lceil\frac{q(h-A)}{m-q}\right\rceil\le A,
\]
again by $qh\le mA$. Multiplying \eqref{pw:missing-ends}
by $K$ gives $KL_0+z^hKH_0=0$. The first term has degree
less than $A+d=h$; the second starts at degree $h$.
They cannot cancel across these two ranges, so both vanish.
Every nonzero solution of $Ky=0$ has degree at least $d$,
whereas $L_0,H_0$ have smaller degree. Thus both are zero.
Finally $C=Z^{\mathsf T}$ has full column rank, so
$CX_*=0$ implies $X_*=0$.

Both cases therefore give a unique tail, which proves the
rank assertion used in the final-block lemma below.
\end{proof}

\subsection{\srnamefix{Decoding the Final SR-2 Block $\mathbf P_{p-1}$}}
\begin{lemma}[Ordinary recovery from the final SR-2 block]
\label{lem:sr2-final-block}
For a burst at $0,\ldots,v-1$, the final block
$\mathbf P_{p-1}$ recovers each $B[i]$ and $A^{\rm o}[i]$
by $E+i$, and each $F[i]$ by $D+i$.
This recovery uses only systematic inputs outside the burst and
does not assume prior knowledge of $U$ innovations.
\end{lemma}
\begin{proof}
Subtract the known outside-burst inputs from each received parity.
The $G_j$ and $U$ identity taps have no erased-input contribution
at times $pv$ and later. We read the three final windows in order.

\noindent\emph{1) The auxiliary window $[pv,E-1]$ and the $B$ group.}
Before $D$, the first $q$ parity coordinates give
$B[i]$ at $pv+i$ for $i<L$.
If $h\le0$ this is the entire burst.
Otherwise Lemma~\ref{pw:tail-rank} gives the tail by $E-1$
when $m\ge q$, or by $D-1$ after prefix subtraction when $m<q$.
For a tail index $i\ge L=A+q$,
$D-1=E+q-1\le E+i$. The prefix deadline $pv+i\le E+i$
also holds, so every $B[i]$ has ordinary delay at most $E$.

\noindent\emph{2) The $A^{\rm o}$ identity, starting at $E$.}
The last $m$ coordinates at $E+i$ contain $A^{\rm o}[i]$
and auxiliary $B$ terms. For the direct and pure-tail assignments,
$\lambda_j\le pv-L$; thus any interfering erased $B$ index
is at least $A+L+i\ge L$. These tail packets are known by
$E-1$. For the prefix-subtracted assignment,
$\lambda_j\le\delta+A=E-v$, so the interfering indices
are at least $v+i$ and their systematic packets are known.
If $h\le0$ there is no auxiliary term. Hence subtraction reveals
$A^{\rm o}[i]$ by $E+i$.

\noindent\emph{3) The final $F$ identity, starting at $D$.}
The first $q$ coordinates at $D+i$ give $F[i]+B[L+i]$.
The second term is outside the burst or belongs to the tail known
by $D-1$. Subtract it to recover $F[i]$ by $D+i$.
This proves all three ordinary deadlines without using an extended
innovation as side information.
\end{proof}

\subsection{Proof of the SR-2 Delay Profile}
\label{app:sr2-profile-proof}
\begin{proof}[Proof of Lemma~\ref{pw:sr2-profile-lemma}]
Fix a burst at $0,\ldots,v-1$. We follow the column blocks of
\eqref{pw:sr2-blocks}, keeping older $B$ terms in the early mixture.

\noindent\emph{1) Blocks $\mathbf P_0,\ldots,\mathbf P_{p-3}$.}
For $1\le j\le p-2$, the identity tap at $jv$ gives
$G_j[i]$ at $jv+i$. Other source groups have no erased-input
contribution in that window. The delay is $jv$.
These blocks are absent when $p=2$.

\noindent\emph{2) Block $\mathbf P_{p-2}$.}
At time $\delta+i$, the $U$ identity and the $B$ coupling block
$\mathcal B_U$ give $U[i]+M[i]$ after subtracting the other
known terms. In particular, the main $B$ tap contributes
$B[i-v]J_F$ and an auxiliary tap contributes
$B[i-(\theta-j)]C_j^{\mathsf T}J_A$.
If an outside-burst contribution was removed while cleaning the
parity, add it back to form the prescribed full innovation.
All nonzero correction lags lie in $[1:v]$, so they are strictly
positive. Thus $I_U[i]$ is known by $\delta+i$.

\noindent\emph{3) Block $\mathbf P_{p-1}$.}
Lemma~\ref{lem:sr2-final-block} gives ordinary delays $E,E,D$
for $B,A^{\rm o},F$. Its proof is independent of Step 2.

\begin{refineblock}
\noindent\emph{4) The innovation is available at every source time.}
Put $z=t+\delta$. If the parity packet at $z$ is received,
subtract $F[z-D]$, $A^{\rm o}[z-E]$, and $G_j[z-jv]$ in
\eqref{pw:source-parities}. Their ordinary deadlines are at
most $z$, by Steps 1 and 3, and the remainder is exactly $I_U[t]$.
If that packet is erased in a burst starting at $c$, then
$\delta\ge v$ gives $t=z-\delta<c$.
Every symbol in $I_U[t]$ has source time at most $t$ and
therefore precedes this isolated burst. Its value is known
from systematic packets. This argument uses only the ordinary
recoveries already proved, for every source time.
\end{refineblock}

Only $U$ is transformed, and its correction uses strictly older
symbols of the ordinary $B$ group. The profile is therefore
separable. The older $B$ symbols are all known by $t+E-1$,
so $U[t]$ itself is also known by $t+D$.
Time translation proves the result for any isolated burst start.
Shorter bursts can be padded up to the deadline being checked.
\refine{In the component's $(v,D)$ sliding-window model,
the next burst begins after every erased input of the preceding
burst has met its $D$-deadline. Reconstructing those inputs
therefore reduces each successive burst to the isolated case.
This also supplies any older inputs needed in Step 4.}
\end{proof}
\subsection{\srnamefix{SR-2 Relay Maps and End-to-End Recovery}}\label{app:sr2-pairing}

\srnamefix{For $(b_1,b_2)=(v,u)$, place SR-2 on the SR link
and the common ordinary code of Lemma~\ref{RD} on the RD link, with $B=u$.}
The relay forms
\begin{equation}
 \begin{gathered}
 R_F[t]=F[t-D],\quad
 R_A[t]=A^{\rm o}[t-E],\quad R_B[t]=B[t-E],\\
 R_U[t]=I_U[t-\delta],\qquad
 R_{G_j}[t]=G_j[t-jv].
 \end{gathered}
 \label{pw:forward-map}
\end{equation}
\srnamefix{The SR-2 extended profile of Lemma~\ref{pw:sr2-profile-lemma}
makes every entry available at time $t$.}
The end-to-end delays are
\begin{equation}
 \begin{array}{c|c|c|c|c}
 \text{group}&\text{SR link delay}&\text{RD link delay}&\text{sum}
                                      &\text{delivered symbol}\\ \hline
 F&D&u&T&F[t]\\
 A^{\rm o},B&E&u+q&T&A^{\rm o}[t],B[t]\\
 U&\delta&2u+q&T-(p-2)(u-v)&I_U[t]\\
 G_j&jv&(p-j+1)u+q&T-(j-1)(u-v)&G_j[t]
 \end{array}
 \label{pw:delay-pairing}
\end{equation}
The destination obtains $I_U[t]$ by $t+T$. Each $B[t-s]$
in $M[t]$ has $s\ge1$ and is delivered individually by
$t-s+T<t+T$. It can therefore subtract $M[t]$ and recover
the original $U[t]$ by $t+T$.

\srnamefix{For order $(u,v)$, place the common ordinary code of
Lemma~\ref{RD} on the SR link and SR-2 on the RD link.} The relay receives
ordinary source symbols with the profile of Lemma~\ref{RD},
using the group identification in Section~\ref{sec:SR2construction}.
It compensates for the second hop's correction by setting
\begin{equation}
 \begin{gathered}
 R_F[t]=F[t-u],\quad
 R_A[t]=A^{\rm o}[t-u-q],\quad R_B[t]=B[t-u-q],\\
 R_{G_j}[t]=G_j[t-(p-j+1)u-q],\\
 R_U[t]=U[t-2u-q]+M[t-u-q].
 \end{gathered}
 \label{pw:reverse-map}
\end{equation}
\srnamefix{The $U$ term is available by its ordinary decoding deadline on the SR link.}
Every $B$ term of $M[t-u-q]$ has a time index strictly
before $t-u-q$, so it too is already available at the relay.
The second hop recovers, with delay $\delta$, the extended
symbol
\begin{equation}
 \begin{split}
 I_{R_U}[t]
 &=R_U[t]+R_B[t-v]J_F
       +\sum_{j=0}^{d}R_B[t-(\theta-j)]C_j^{\mathsf T}J_A\\
 &=U[t-2u-q]+M[t-u-q]+M[t-u-q]\\
 &=U[t-2u-q].
 \end{split}
 \label{pw:cancellation}
\end{equation}
Thus the added term in \eqref{pw:reverse-map} cancels exactly
the correction term of the RD extended symbol. The original
source groups now have delay sums
\begin{equation}
 \begin{array}{c|c|c|c}
 \text{group}&\text{SR link delay}&\text{RD link delay}&\text{sum}\\ \hline
 F&u&D&T\\
 A^{\rm o},B&u+q&E&T\\
 U&2u+q&\delta&T-(p-2)(u-v)\\
 G_j&(p-j+1)u+q&jv&T-(j-1)(u-v)
 \end{array}.
 \label{pw:reverse-delays}
\end{equation}
For $U$, the delay $\delta$ in this table follows from the
cancellation in \eqref{pw:cancellation}. The same two component
encoders are used in reverse order.

Both relay maps are fixed linear expressions in the source stream.
The erasure pattern affects how the relay computes their values,
but not the values it transmits.

Both components have $(k,n)=(T-v,T-v+u)$.
The fixed relay maps and the verified delay sums prove
Theorem~\ref{pw:achievability}.

\begin{revisionblock}
\subsection{\srnamefix{An SR-2 Zigzag Example Beyond the SR Range}}
\label{ex:zigzag-new-range}
This example uses two received packets to recover fifteen tail symbols.
Take $(v,u,T)=(10,12,37)$. Then
\[
 (p,q)=(2,3),\quad (k,n)=(27,39),\quad
 (m,A,L,h)=(9,2,5,5),\quad(\delta,E,D)=(10,22,25).
\]
The SR condition $q\le A$ fails, whereas the SR-2 condition
holds: $qh=15\le mA=18$.
The recovery below exhibits the full row rank of the finite tail matrix directly.

\noindent\emph{\srnamefix{1) The SR-2 coefficient assignment.}}
Use source groups $(F,A^{\rm o},B,U)$ of sizes $(3,9,3,12)$.
There are no $G_j$ groups. The pure-tail choice in
\eqref{pw:coefficient-choice} is
$C(z)=\ZZ(3,6,3)^{\mathsf T}$, namely
\begin{equation}
 C(z)=
 \begin{pmatrix}
 \nz1&0&0\\
 \nz{z^2}&0&0\\
 \nz{z^3}&\nz1&0\\
 0&\nz z&0\\
 0&\nz{z^3}&0\\
 0&0&\nz1\\
 0&0&\nz z\\
 0&0&\nz{z^2}\\
 0&0&\nz{z^3}
 \end{pmatrix}
 =\sum_{j=0}^3 C_jz^j,
 \qquad \theta=5,\quad \lambda_j=15-j.
 \label{eq:example-zigzag-C}
\end{equation}
\fourthfix{Here $C(z)$ has nine rows and three columns.
Its sixth row is $(0,0,1)$, supplied by $(C_0)_{6,3}=1$.
All other coefficients in that row are zero.}
Thus the source sends $(S[t],P_F[t],P_A[t])$, where
\begin{equation}
 \begin{aligned}
 P_F[t]&=F[t-25]+B[t-20]+U_F[t-10],\\
 P_A[t]&=A^{\rm o}[t-22]+U_A[t-10]
                  +\sum_{j=0}^3 B[t-15+j]C_j^{\mathsf T}.
 \end{aligned}
 \label{eq:example-zigzag-parity}
\end{equation}
All sums are over $\mathbb F_2$. On the longer-burst hop, use
Lemma~\ref{RD} with burst length $12$, head size $3$, and
identity taps at $15$ and $27$.

\noindent\emph{2) Read the fifteen tail symbols from two packets.}
\srnamefix{For order $(10,12)$, erase packets $0,\ldots,9$ on the SR link.} The first three parity positions
at times $20,\ldots,24$ give the prefix $B[0],\ldots,B[4]$.
For the tail, use the last nine parity positions at times $20,21$.
Subtract their known systematic contributions and denote the results
by $y_0,y_1$. Writing $B=(B_1,B_2,B_3)$, the observations are
\begin{equation}
 \begin{array}{c|c|c}
 \text{position}&y_0\ \text{(time 20)}&y_1\ \text{(time 21)}\\ \hline
 1&B_1[5]&B_1[6]\\
 2&B_1[7]&B_1[8]\\
 3&B_1[8]+B_2[5]&B_1[9]+B_2[6]\\
 4&B_2[6]&B_2[7]\\
 5&B_2[8]&B_2[9]\\
 6&B_3[5]&B_3[6]\\
 7&B_3[6]&B_3[7]\\
 8&B_3[7]&B_3[8]\\
 9&B_3[8]&B_3[9]
 \end{array}.
 \label{eq:example-zigzag-observations}
\end{equation}
Read the unmixed entries first. The two remaining symbols are
\[
 B_2[5]=y_{0,3}+B_1[8],\qquad
 B_1[9]=y_{1,3}+B_2[6].
\]
Hence all of $B[5],\ldots,B[9]$ is known at time
$21=E-1$, so the two auxiliary packets recover all fifteen tail symbols.
The $15\times18$ tail matrix therefore has row rank $15$.
The monomial terms in \eqref{eq:example-zigzag-C} spread the five tail packets
across the two auxiliary packets.

\noindent\emph{\srnamefix{3) Complete SR-2 decoding and pair the components.}}
\srnamefix{The SR-2 decoding order now specializes to a short sequence.}
The prefix and tail give every $B[i]$ by $i+22$.
At $i+22$, remove the now available $B$ terms from $P_A$ to
recover $A^{\rm o}[i]$. At $i+25$, remove $B[i+5]$ from
$P_F$ to recover $F[i]$; it is either in the recovered tail or
outside the burst. The early $U$ window supplies
\begin{equation}
 I_U[t]=\biggl(U_F[t]+B[t-10],\quad
 U_A[t]+\sum_{j=0}^3B[t-5+j]C_j^{\mathsf T}\biggr)
        =U[t]+M[t]
 \label{eq:example-zigzag-innovation}
\end{equation}
by $t+10$. \fourthfix{Here $t$ is the source time, and $t+10$
is the channel time at which its innovation is due.
The auxiliary terms in $P_A[t+10]$ have indices
$t+10-15+j=t-5+j$, exactly as in
\eqref{eq:example-zigzag-innovation}.}
Lemma~\ref{pw:sr2-profile-lemma} also guarantees this innovation
at unerased source times, including when $S[t]$ is received
systematically.
The coordinate delays pair as follows:
\[
 \begin{array}{c|c|c|c}
 \text{group}&\text{SR-2 component}&
       \text{common ordinary component}&\text{sum}\\ \hline
 F&25&12&37\\
 A^{\rm o},B&22&15&37\\
 U&10\ \text{(extended)}&27&37
 \end{array}.
\]
For order $(10,12)$, the relay input is
\[
 R[t]=(F[t-25],A^{\rm o}[t-22],B[t-22],I_U[t-10]).
\]
The destination gets $I_U[t]$ by $t+37$; all $B$ terms in
$M[t]$ have source times at most $t-2$ and have already arrived,
so it recovers $U[t]$. For order $(12,10)$, use
\[
 R[t]=(F[t-12],A^{\rm o}[t-15],B[t-15],
                         U[t-27]+M[t-15]).
\]
Each term is known at the relay: the $B$ terms in $M[t-15]$
are strictly older than $t-15$ and have ordinary delay $15$.
The second hop's $U$ innovation adds another copy of $M[t-15]$,
leaving $U[t-27]$ after delay $10$. Thus both link orders recover
every original source packet within $37$, at the rate upper bound
$27/39=9/13$.

\end{revisionblock}

\section{\eighthfix{Supporting Proofs and Point-to-Point Lemmas}}
\label{app:ordinary-proofs}
\subsection{The SR Transition Block}\label{app:sr-transition}
\begin{proof}
For $q<v$, the retained matrix is exactly the head matrix
$H_0$ in Lemma~\ref{invertible}, with its parameter $B=v$,
shifted by $\delta$. Its identity tap is at $\delta+v=pv$,
and its sparse taps are $\delta+s$ with coefficients $\fourthfix{\mathsf{L}_s}$.
The lemma gives the individual deadlines $pv+i$ and full
recovery by $pv+q-1=H-1$. For $q\ge v$, the sparse taps
are zero and $pv+i<H$ for every erased index $i<v$.
The first $q$ main parity positions therefore give $B[i]$
without interference from $V$.
\end{proof}
\subsection{The Layered Component Family}\label{app:fixed-family}
\begin{proof}
For $q>0$, the $v$-burst layered component assigns delay
$(p+1)v$ to its $q$ additional coordinates. Every coordinate on
the other hop needs at least delay $u$. No permutation can evade
this minimum. Hence a valid pairing
requires $(p+1)v+u\leq T$.
The reverse-order pairing in the large-delay construction attains
this requirement: the additional coordinates have delay sum
$(p+1)v+u$, and the other blocks have sums
$(p-j)v+(j+1)u+q\leq T$, $0\leq j<p$.

For $q=0$, there is no additional block. The divisible
construction pairs delays $(p-j)v$ and $(j+1)u$, whose sums
are at most $T=v+pu$. Thus the ceiling form covers both cases.
\end{proof}
\subsection{Supporting Point-to-Point Lemmas}
\refine{The following systematic-packet count is a special case
of the two-stream delay tradeoff in~\cite[Theorem~2]{BadrMultiplexed2018}.}
\begin{lemma}[A late-coordinate count]
\label{secvi:late-coordinates}
Consider a systematic point-to-point burst-erasure code with burst
length $B$, dimensions $(k,k+r)$, and source-time-uniform ordinary
coordinate deadlines at most $D>B$.
Let $k_D$ count the coordinates whose minimum guaranteed delay is
$D$. Then
\begin{equation}
 Bk-k_D\le(D-1)r.
 \label{secvi:late-count-bound}
\end{equation}
In particular, if $k/(k+r)=D/(D+B)$, then $k_D\ge r$.
\end{lemma}
\begin{proof}
Take independent uniform input symbols and erase packets
$c,\ldots,c+B-1$. By time $c+D+B-2$, the $k-k_D$ coordinates
due by $D-1$ in all $B$ erased packets and the remaining
coordinates in the first $B-1$ packets must be known.
There are $Bk-k_D$ targets.
Given all outside-burst inputs through this time, only the $D-1$
received parity packets can reveal information about them.
This proves \eqref{secvi:late-count-bound} without linearity or
time invariance. Rate equality gives $Bk=Dr$, hence $k_D\ge r$.
\end{proof}

\begin{lemma}[An early-coordinate count]
\label{secvi:early-count}
Consider a systematic causal point-to-point code with $k$ input
symbols and $r$ parity symbols per packet. It corrects a burst of
length $b$ and recovers every input packet within delay $b+q$,
where $0<q<b$. If $m$ input coordinates per packet are recoverable
within delay $b$, then
\begin{equation}
 k+qm\leq(q+1)r.
 \label{secvi:early-count-bound}
\end{equation}
\end{lemma}
\begin{proof}
Erase packets $t,\ldots,t+b-1$. By $t+b+q$, the decoder must
recover the full packet at $t$ and the $m$ early coordinates from
each of the packets at $t+1,\ldots,t+q$. Let $W$ collect these $k+qm$ erased symbols and let $G$
contain all input packets outside the burst through the deadline.
Take independent uniform input symbols over $\mathbb F$.
Systematic observations and preburst packets are known given $G$;
only the $q+1$ received parity packets can contain information about $W$.
Zero-error recovery therefore gives
\[
\begin{split}
(k+qm)\log|\mathbb F|&=H(W\mid G)\\
 &=I(W;P[t+b:t+b+q]\mid G)\\
 &\leq(q+1)r\log|\mathbb F|.
\end{split}
\]
This proof does not require linearity or time invariance.
\end{proof}

\begin{lemma}[A burst-prefix count]
\label{lem:ordinary-prefix-count}
Consider a systematic causal component with burst length $B$,
dimensions $(k,k+r)$, and ordinary coordinate delays
$\delta_1,\ldots,\delta_k$. Let
$F(h)=|\{i:\delta_i\leq h\}|$, with $F(h)=0$ for $h<B$.
Then, for every integer $h\geq B$,
\begin{equation}
 \sum_{\ell=0}^{B-1}F(h-\ell)\leq(h-B+1)r.
 \label{eq:ordinary-prefix-count}
\end{equation}
Linearity and time invariance are not required.
\end{lemma}
\begin{proof}
Erase the $B$ packets at $c,\ldots,c+B-1$ and give the
decoder all outside-burst inputs through time $c+h$.
By that time, at least $F(h-\ell)$ ordinary coordinates of the
erased input at $c+\ell$ must be recovered.
These targets are independent uniform source symbols.
Given the genie information, only the $h-B+1$ received parity
packets can contain information about them, at most
$(h-B+1)r$ symbols in total. The same conditional-entropy
argument as in Lemma~\ref{secvi:early-count} proves the inequality.
\end{proof}

\section{An Ordinary Code Outside the Layered Baseline Range}
\label{app:ordinary-counterexample}
We verify the code stated in Example~\ref{secvi:p3-example}.

\noindent\emph{Design idea: use the spare delay budget.}
The old four-burst layered allocation would assign delay $16$
to the additional coordinate $z$. The other component needs at
least delay $5$, so this allocation exceeds $T=20$.
We instead recover $z$ at delay $15$, while allowing the $C$
group delay $14$ instead of $12$ and the $A$ group delay $9$
instead of $8$; the $U$ group keeps delay $4$.
The five-burst profile is unchanged. Thus the $C$ and $A$ groups
use their spare delay budget, while the bottleneck $z$ moves one
slot earlier, and every paired sum becomes $20$.
The early $C_1$ taps below make this redistribution decodable.
The recovery argument below verifies all deadlines directly.
All additions occur in component parity packets, not the relay
input map; the following equations verify the proposed deadlines.

\begin{center}
\begin{tabular}{c|cccc}
Coordinate group & $z$ & $C_1,\ldots,C_5$ & $A_1,\ldots,A_5$ & $U_1,\ldots,U_5$\\
\hline
Old four-burst delay & $16$ & $12$ & $8$ & $4$\\
New four-burst delay & $15$ & $14$ & $9$ & $4$\\
Five-burst delay & $5$ & $6$ & $11$ & $16$\\
New total delay & $20$ & $20$ & $20$ & $20$
\end{tabular}
\end{center}

Use the input order
\[
 W[t]=(z[t],C_1[t],\ldots,C_5[t],
              A_1[t],\ldots,A_5[t],U_1[t],\ldots,U_5[t]).
\]
All negative-time symbols are zero.
The four-burst component sends $X[t]=(W[t],\eta[t])$, where
\begin{align}
 \eta_1[t]&=U_1[t-4]+A_1[t-8]+C_1[t-11]+z[t-15],\notag\\
 \eta_2[t]&=U_2[t-4]+A_2[t-8]+C_1[t-12]+C_2[t-14],\notag\\
 \eta_j[t]&=U_j[t-4]+A_j[t-8]+C_j[t-14],
                  \quad j=3,4,5.
 \label{secvi:four-burst-encoder}
\end{align}
We claim ordinary delays $15$ for $z$, $14$ for the $C$ group,
$9$ for the $A$ group, and $4$ for the $U$ group.

\noindent\emph{1) Recover the four-burst component.}
Let the erased packets be $0,1,2,3$ and subtract
known contributions from all other source times.
First, every $U_j[i]$ is directly obtained at $i+4$.
The $A$ symbols are directly obtained at $i+8$, except
$A_1[3]$, which appears together with $C_1[0]$ at time $11$.
At time $12$, $\eta_2[12]$ reveals $C_1[0]$, so $A_1[3]$
is recovered at $12=3+9$.
The first parity coordinate reveals $C_1[1],C_1[2],C_1[3]$
at times $12,13,14$. Hence all four erased $C_1$ symbols
are known by $14$.

For $j=2,\ldots,5$, the primary term $C_j[i]$ arrives at
$i+14$. The only possible additional erased term is
$C_1[i+2]$ in the second parity coordinate. If $i=0,1$,
that term has already been recovered; if $i=2,3$, it is
outside the burst and known from the systematic packets.
Thus all $C$ coordinates meet delay $14$.
Finally, $z[i]$ arrives at $i+15$ in the first parity
coordinate. Its other terms are outside the burst, including
$C_1[i+4]$, so it is directly recovered.
Time invariance gives the same argument for any burst start.
Shorter bursts are covered by extending them to a four-packet
burst and ignoring the additional available packets.
In the original sliding-window model, inputs erased in a previous
burst are recovered before the next burst starts. The same
known-past subtraction therefore applies to successive bursts.

\noindent\emph{2) Pair with the unchanged five-burst component.}
The five-burst component is the ordinary RD construction with
$p=3,q=1$. For clarity, its parity equations are
\begin{align}
 \rho_1[t]&=z[t-5]+C_1[t-6]+A_1[t-11]+U_1[t-16],\notag\\
 \rho_j[t]&=z[t-(j-1)]+C_j[t-6]+A_j[t-11]+U_j[t-16],
                   \quad j=2,\ldots,5.
 \label{secvi:five-burst-encoder}
\end{align}
Here the symbols denote this component's own input coordinates;
on the RD hop they are coordinates of $R[t]$.
Under a burst at $0,\ldots,4$, the parity packet at time $5$
recovers all five erased $z$ symbols. Then the $C$, $A$ and
$U$ groups are directly recovered at delays $6$, $11$ and
$16$, respectively, after known terms are subtracted.
The two profiles pair as follows:

For the order $(b_1,b_2)=(4,5)$, the relay uses exactly
\begin{equation}
 R[t]=(z[t-15],C_1[t-14],\ldots,C_5[t-14],
             A_1[t-9],\ldots,A_5[t-9],
             U_1[t-4],\ldots,U_5[t-4]).
 \label{secvi:p3-relay}
\end{equation}
The coordinate delay sums are $15+5=14+6=9+11=4+16=20$.
Every forwarded coordinate is an original source symbol; no
extended innovation or addition appears in the relay map.
Interchanging the two components gives the same rate and deadline
for $(5,4,20)$.

The old additional coordinate needs delay $16$ and cannot be
paired with any RD coordinate of delay below $5$.
Reordering the old profiles alone therefore cannot meet $T=20$.
The new four-burst encoder changes the delay distribution, making
every new sum $20$.
This lies outside the low-delay regime of Theorem~\ref{secvi:theorem20}
and the equal-burst regime of Theorem~\ref{thm:equal-ordinary-allp}. It shows exactly why a necessary condition for the
old fixed family cannot be promoted to all ordinary codes.

\subsection{A Partial Infinite Family for $p=3$}

\begin{lemma}[Time-interleaving closure]
\label{lem:time-interleaving}
A fixed-rate code for $(b_1,b_2,T)$ yields a code of the same
packet dimensions and rate for
$(\lambda b_1,\lambda b_2,\lambda T)$, for every integer
$\lambda\geq1$. Ordinary coordinate forwarding, binary linear
components, and fixed extended innovations are each preserved.
\end{lemma}
\begin{proof}
Run $\lambda$ independent copies on the time classes modulo
$\lambda$. Replace each component tap, innovation lag, and relay
shift by $\lambda$ times its original value.
A global burst of at most $\lambda b_i$ packets intersects
each time class in at most $b_i$ consecutive local packets.
A local window of $T+1$ instants spans $\lambda T+1$ global
instants, so the required burst guarantee is preserved.
Every local delay $d$ becomes $\lambda d$, with no change in
symbols per packet. This is interleaving separate streams, not
merging packets into a different channel model.
\end{proof}

\begin{corollary}[An ordinary-forwarding $p=3$ family]
\label{cor:p3-family}
For every integer $\lambda\geq1$, binary ordinary forwarding
attains $R^\star=16/21$ at
$(4\lambda,5\lambda,20\lambda)$ and its swapped ordering,
with $(k,n)=(16,21)$.
All these parameters fail the layered baseline condition.
\end{corollary}
\begin{proof}
Apply Lemma~\ref{lem:time-interleaving} to
Example~\ref{secvi:p3-example}. Here
$T-v=16\lambda=3(5\lambda)+\lambda$, whereas
$T-u=15\lambda<4\lambda\lceil16/5\rceil=16\lambda$.
\end{proof}
This extends the counterexample along a scaled parameter ray.
The complete unequal-burst ordinary-forwarding region for $p\geq3$
is left open; no such characterization is used in the proved results.

\bibliographystyle{IEEEtran}
\bibliography{assets/references}
\end{document}